\documentclass[prd,showpacs,nofootinbib,onecolumn,amsmath,amsfonts,amssymb]{revtex4} %,nofootinbib

\usepackage{amsmath}
\usepackage{amssymb}
\usepackage{epsfig}

\usepackage{graphicx}
\usepackage{stmaryrd}
\usepackage{bm}
\usepackage{color}

\newcommand{\eps}  {\epsilon}

\newcommand{\AddrAHEP}{%
 Cluster of Excellence, Origin and Structure of the Universe, Technische Universit\"{a}t M\"{u}nchen,\\
 Boltzmannstra\ss e 2, D-85748, Garching, Germany}

\begin{document}

\title{Testing the very-short-baseline neutrino anomalies at the solar sector}

\author{Antonio Palazzo}

\affiliation{\AddrAHEP}

\date{\today}

\begin{abstract}
 
  Motivated by the accumulating hints of new sterile neutrino species at the eV scale, we explore
  the consequences of such an hypothesis on the solar sector 
  phenomenology. After introducing the theoretical formalism needed to describe the MSW conversion 
  of solar neutrinos in the presence of one (or more) sterile neutrino state(s) located 
  ``far'' from the ($\nu_{1}$, $\nu_{2}$) ``doublet'', we perform 
 a quantitative analysis of the available experimental results, focusing on the electron neutrino
 mixing.  We find that the present data posses 
  a sensitivity to the amplitude of the lepton mixing matrix element $U_{e4}$ ---  encoding the 
  admixture of the electron neutrino with a new mass eigenstate --- which is comparable to that achieved 
  on the standard matrix element $U_{e3}$. In addition, and more importantly,  our analysis 
  evidences that, in a 4-flavor framework, the current preference for $|U_{e3}|\ne 0$ is indistinguishable
  from that for $|U_{e4}|\ne0$, having both a similar statistical significance (which is $\sim 1.3 \sigma$ 
  adopting the old reactor fluxes determinations, and $\sim 1.8\sigma$ using their new estimates.) 
  We also point out that, differently from the standard 3-flavor case, in a 3+1 scheme the Dirac CP-violating
  phases cannot be eliminated from the description of solar neutrino conversions.

\end{abstract}


\maketitle

\section{Introduction}

The interest around the possible manifestation of novel neutrino  properties
in very short baseline (VSBL)  setups has been recently reawakened by the
emergence of new (and the reappraisal of old) inconsistencies.  Improved calculations of the reactor antineutrino spectra~\cite{Mueller:2011nm} suggest 
fluxes which are $\sim 3\%$ higher than previous estimates~\cite{Vogel:1980bk, VonFeilitzsch:1982jw,Schreckenbach:1985ep,Hahn:1989zr},
and have raised the so-called antineutrino anomaly~\cite{Mention:2011rk}, consisting in a deficit
of (almost) all the VSBL reactor measurements performed at distances $L\lesssim 100 $~m.
An apparently unrelated deficit had been already evidenced in the calibration measurements 
conducted at the solar neutrino experiments GALLEX and 
SAGE~\cite{Abdurashitov:2005tb,Giunti:2010zu}, employing radioactive sources 
placed inside the detectors. Such two discrepancies join  
those already recorded at the VSBL accelerator experiments~\cite{Aguilar:2001ty, AguilarArevalo:2007it, AguilarArevalo:2009xn}, adding further confusion to 
an already  intricate ``VSBL neutrino puzzle''.

While some of the (old and new) anomalies may be imputable to theoretical and/or experimental 
inaccuracies, the possibility exists that they may represent a manifestation of new physics.
From this perspective, they seem to point towards a phenomenon of (dis-)appearance of the electron
neutrinos, possibly mediated by oscillations into new sterile specie(s)~\cite{Giunti:2010zu,Mention:2011rk}.  
Such an hypothesis finds support in the lesser degree of tension  --- although still a considerable one ---
now existing between the (weakened)  VSBL  reactor limits  and the (unchanged) VSBL accelerator constraints, 
within all schemes endowed with additional sterile species~\cite{Giunti:2010uj, Kopp:2011qd}.
Notably, an independent hint in the same direction arises from the latest cosmological data
analyses~\cite{Hamann:2010bk}, which favor a significant extra relativistic energy content,  although pointing
towards masses of the new presumptive light particles~\cite{Hamann:2010bk,Giusarma:2011ex}, which are at the borderline of those (larger) suggested for the sterile neutrinos by the oscillation data.%
%%%%%%%%%%%%%%%%%%%%%%%%%%%%%%%%%%%%%%%%%%%%%%%%%%%%
\footnote{It must be stressed that the degeneracy between the neutrino masses and the dark energy
equation of state parameter $w$~\cite{Hannestad:2005gj} may allow the accommodation of such bigger
masses at the cost of assuming $w<-1$ and a lower expansion age of the universe,
as recently remarked in~\cite{Kristiansen:2011mp}.}
%%%%%%%%%%%%%%%%%%%%%%%%%%%%%%%%%%%%%%%%%%%%%%%%%%%%
Such a trend is partially corroborated by the latest  constraints coming from primordial
nucleosynthesis~\cite{Izotov:2010ca, Mangano:2011ar},  which can easily accommodate one, but hardly 
two additional sterile species. 

The putative sterile neutrinos must be introduced  without spoiling the basic success  of the 
standard 3-flavor paradigm. This can be achieved in the so-called $3+s$ schemes, where the 
$s$ new  mass eigenstates are assumed to be separated from the three standard ones by large splittings,%
%%%%%%%%%%%%%%%%%%%%%%%%%%%%%%%%%%%%%%%%%%%%%%%%%%%%
\footnote{Another (possibly coexisting) realization is provided by new sterile species separated
from the active ones by extremely small mass-squared splittings~\cite{deHolanda:2003tx,  deHolanda:2010am}. However, such schemes do not have observable effects in VSBL settings and
we do not consider them in this work.}
%%%%%%%%%%%%%%%%%%%%%%%%%%%%%%%%%%%%%%%%%%%%%%%%%%%%
giving rise to the hierarchal pattern%
\footnote{The solar data are insensitive to the reciprocal ordering of the third
and fourth mass eigenstates, as well as  to their ordering respect to the ``doublet'' $(\nu_1,\nu_2)$.
In this work we assume for definiteness  $m_1 < m_2 <m_3<m_4$,
always referring to the positive mass-squared splittings defined as $\Delta m^2_{ij} = m^2_j - m^2_i$ ($i<j$).}
$|m^2_2 - m^2_1 | \ll  |m^2_3 - m^2_1 | \ll |m^2_k - m^2_1 |\,\, (k = 4, ..., 3+s)$,
%%%%%%%%%%%%%%%%%%%%%%%%%%%%%%%%%%%%%%%%%%%%%%%%%%%%
which ensures that the fast oscillations induced by the new mass eigenstates
are completely averaged in all settings sensitive to the $\Delta m^2_{12}$-driven (solar) and 
$\Delta m^2_{13}$-driven (atmospheric) transitions, leaving unmodified the two well-established oscillation frequencies.
With the additional assumption of a small admixture of
the active flavors with the new mass eigenstates, the $3+s$ schemes
leave basically unaltered also the standard oscillation amplitudes,
thus realizing a genuine perturbation of the leading 
3-flavor scenario,  whose size must respect the constraints imposed by 
all the existing phenomenology.%
%%%%%%%%%%%%%%%%%%%%%%%%%%%%%%%%%%%%%%%%%%%%%%%%%%%%
\footnote{This includes the atmospheric neutrino data collected at neutrino telescopes~\cite{Abbasi:2010ie}, where the
new sterile species may leave distinctive imprints~\cite{Nunokawa:2003ep,Choubey:2007ji,Razzaque:2011ab}.}
%%%%%%%%%%%%%%%%%%%%%%%%%%%%%%%%%%%%%%%%%%%%%%%%%%%%

The solar sector data (Solar and KamLAND) have had a pivotal
role in establishing and shaping the 3-flavor framework,
and continue to be of extreme importance in sharpening its basic parameters.
As a matter of fact, these data are the only ones sensitive to the admixture
of the electron neutrino with the first two mass eigenstates ($\nu_1, \nu_2$),
also possessing  a subleading sensitivity to the (averaged) $\nu_3$-driven  oscillations.
This translates into stringent constraints on the amplitude of the elements ($U_{e1}, U_{e2}, U_{e3}$) 
entering the first row of the lepton mixing matrix. Therefore,
it is of certain interest to explore how such a sector ``responds'' 
to the perturbations generated by  a non-zero mixing of the electron neutrino
with new sterile species, as hinted at by the recent VSBL findings.

The impact of new sterile species on the phenomenology of the solar neutrino sector
has been investigated in several works, both as a leading mechanism~\cite{Barger:1990bg,Krastev:1996gc}
(in the ``pre-KamLAND era'') and  as a subdominant one~\cite{Giunti:2000wt,Bahcall:2002zh,deHolanda:2002ma,Bahcall:2002ij,Maltoni:2002ni,Cirelli:2004cz,GonzalezGarcia:2007ib} (after the KamLAND results). 
However, all the existing  analyses% 
%%%%%%%%%%%%%%%%%%%%%%%%%%%%%%%%%%%%%%%%%%%%%%%%%%%%
\footnote{An exception is constituted by the work~\cite{deHolanda:2002ma}, 
where a qualitative discussion of the impact of non-zero $U_{e3}$ is provided.}
%%%%%%%%%%%%%%%%%%%%%%%%%%%%%%%%%%%%%%%%%%%%%%%%%%%%
have been performed in the simplified framework --- whose formalism was originally
developed in~\cite{Dooling:1999sg} --- of pure ($\nu_1$-$\nu_2$)-driven oscillations, 
which neglects the possible mixing of the electron neutrino with the third standard
mass eigenstate ($U_{e3}=0$) and with a new fourth one  ($U_{e4}=0$). 
In the past, both assumptions were
justified by the limited sensitivity of this dataset, and by the
strong upper bounds put on these matrix elements by the reactor experiments performed
with short~\cite{Apollonio:2002gd} (sensitive to $U_{e3}$) and very short~\cite{Declais:1994su}
 (sensitive to $U_{e4}$) baselines.
Moreover, no hint of transitions into sterile states was evidenced in the aforementioned analyses, 
so there was no reason to extend them beyond  such a simple scheme. 
However, this situation has gradually changed in the recent years.
On the one hand, we have witnessed a substantial increase in the sensitivity 
of the solar sector to possible departures from the simple 2-flavor approximation;
the recent indication of $\theta_{13}>0$~\cite{Fogli:2008jx}
and the possible hint of non-standard MSW dynamics~\cite{Palazzo:2011vg}
(see also~\cite{Palazzo:2009rb})
suggested by these data testify such a new trend.
On the other hand, the new anomalies directly point towards a
relatively big amplitude of $U_{e4}$, which should be now testable
at the solar sector.

The incorporation of the $3+s$ schemes in the description of solar neutrino conversions
is not a trivial task, however, as it requires the treatment of the MSW effect
in the presence of sterile species.
This problem has been recently addressed in~\cite{Giunti:2009xz}, where a parameterization 
 independent form of the lepton mixing matrix has been exploited,
and selected numerical examples have been given for the relevant transition probabilities.
In this work we make a step forward and, by adopting a convenient parameterization of the 
mixing matrix, we put quantitative constraints on these schemes, focusing  on the
electron neutrino mixing. The rest of our paper is organized as follows. 
In Sec.~II we introduce the $3+1$ neutrino framework and present the basic  formulae 
needed to interpret the flavor oscillations in vacuum and in  (solar) matter within such a scheme. 
In Sec.~III we discuss the results of the numerical analysis, drawing our conclusions in Sec.~IV.
Four appendices address the following (more technical) issues: A) The treatment of the 
solar MSW effect in a $3+1$ scheme; B)  Its generalization to the $3+s$ frameworks; 
C) The incorporation of Earth-induced matter effects; D) The potential sensitivity of solar neutrino
flavor transitions to the Dirac CP-violating phases entering the 4-lepton mixing matrix.

\section{Framework and basic analytical results}

\subsection{Parameterization of the  mixing matrix}

In the presence of a fourth sterile neutrino $\nu_s$, the flavor ($\nu_\alpha$, $\alpha = e, \mu, \tau, s$)
and the mass eigenstates ($\nu_i, i =1,2,3,4$), are connected through a $4\times4$
unitary mixing matrix $U$, which depends on six complex parameters~\cite{Schechter:1980gr}.
Such a matrix can thus be expressed as the product  of six complex elementary rotations,  which define six real
mixing angles and six CP-violating phases. Of the six phases three are of the 
Majorana type and are unobservable in oscillation processes, while the three remaining ones
are of the Dirac type. For simplicity, in this work, we set to zero all the  Dirac phases, commenting 
only in Appendix~\ref{sec:app_phases} on the potential sensitivity of the solar data to them.  

As it will appear clear in what follows, for the treatment  of the solar MSW  transitions under study, it is convenient
 to parameterize the mixing matrix as
%..........................................................................
\begin{equation}
\label{eq:U}
U =   R_{23}  R_{24} R_{34} R_{14} R_{13} R_{12}  \equiv B R_{14} R_{13} R_{12}  \equiv A R_{12}\,, 
\end{equation} 
%..........................................................................
where $R_{ij}$ represents a real $4\times4$ rotation in the ($i,j$) plane
containing the $2\times2$ submatrix 
%..........................................................................
\begin{equation}
\label{eq:R_ij_2dim}
     R^{2\times2}_{ij} =
    \begin{pmatrix}
         c_{ij} &  s_{ij}  \\
         - s_{ij}  &  c_{ij}
    \end{pmatrix}
\,,    
\end{equation}
%..........................................................................
in the  $(i,j)$ subblock, with ($c_{ij} \equiv \cos \theta_{ij}, s_{ij} \equiv \sin \theta_{ij}$),
and the matrices $A \equiv UR_{12}^T$ and $B \equiv R_{23} R_{24} R_{34}$
have been introduced  for later convenience.

The  parameterization in Eq.~(\ref{eq:U}) has the following properties: I) For vanishing mixing
involving the fourth state $(\theta_{14} = \theta_{24} = \theta_{34} =0)$ 
Eq.~(\ref{eq:U}) reduces to the 3-flavor mixing matrix in its standard parameterization~\cite{PDG}; 
II) The leftmost positioning of the matrix $R_{23}$ allows us to rotate away the 
mixing angle $\theta_{23}$ from the solar MSW dynamics (see Appendix \ref{sec:app_msw})
and from the expression of the mixing elements involving the sterile flavor (see the discussion below);
III) The positioning of the product $R_{14}R_{13}$ close to the rightmost matrix  $R_{12}$ makes the 
corresponding mixing angles appear in a symmetrical way in the expressions of the 
admixtures of the electron neutrino with the $(\nu_1,\nu_2)$ mass eigenstates, 
inducing small admixtures with the ``far''  $(\nu_3,\nu_4)$ mass eigenstates,
which, in the limit of small $\theta_{14}$ also appear in a  symmetrical form,
being $U_{e3}^2 \simeq \sin^2\theta_{13}$ and $U_{e4}^2 = \sin^2\theta_{14}$. 
In fact, the mixing matrix elements involving the electron neutrino are
expressed as
% .................................................................................................................
\begin{align}
    \label{eq:Ue1_gen} U_{e1} 
     &= A_{11}\, c_{12} - A_{12}\, s_{12} \,, \\\
\label{eq:Ue2_gen} U_{e2}
        &=  A_{11}\, s_{12} + A_{12}\, c_{12} \,,    \\
\label{eq:Ue3_gen} U_{e3} 
        & = A_{13}\,,  \\
\label{eq:Ue4_gen}  U_{e4}
        &= A_{14} \,,
\end{align}
% .................................................................................................................
in terms of the elements of the first row of the matrix A.  The rightmost positioning 
of the product $R_{14} R_{13} R_{12}$ in such a matrix ensures that its
element $A_{12} $ is equal to zero,  thus leading
to the explicit expressions
% .................................................................................................................
\begin{align}
    \label{eq:Ue1} U_{e1} 
     &= c_{14}  c_{13} c_{12}\,, \\\
\label{eq:Ue2} U_{e2}
        &=  c_{14}  c_{13} s_{12}\,,    \\
\label{eq:Ue3} U_{e3} 
        & = c_{14}  s_{13}\,,  \\
\label{eq:Ue4}  U_{e4}
        &= s_{14} \,.
\end{align}
% .................................................................................................................
It should be noted that these expressions are 
not affected by the pre-multiplication of the matrix $B$ in Eq.~(\ref{eq:U}), 
since this induces a rotation in the hyperplane orthogonal to the axis with index $ i=1$,  and therefore
are independent of the specific order of the three matrices ($R_{23}$, $R_{24}$, $R_{34}$).  
Since the solar data are sensitive also to transitions into sterile neutrinos, we will 
need the expressions of the matrix elements involving the sterile flavor, which, taking
into account the last equality in~Eq.~(\ref{eq:U}), can be written as
% .................................................................................................................
\begin{align}
    \label{eq:Us1} U_{s1} 
     &= A_{41} \, c_{12} - A_{42} \, s_{12}\,, \\\
\label{eq:Us2} U_{s2}
        &=   A_{41} \, s_{12}  + A_{42} \, c_{12}\,,  \\
\label{eq:Us3} U_{s3}
        & = A_{43}\,, \\
\label{eq:Us4}  U_{s4}
        &= A_{44} \,,
\end{align}
% .................................................................................................................
in terms of the elements of the fourth row of the matrix A, which, 
in our parameterization have the explicit expressions
%..............................................................................
\begin{align}
   \label{eq:A41} A_{41}
        & =  c_{24} (s_{34}  s_{13} - c_{34} s_{14} c_{13}) \,, \\
 \label{eq:A42} A_{42}
       & = - s_{24}\,,   \\
\label{eq:A43} A_{43}
      & =  - c_{24} (s_{34} c_{13}  + c_{34} s_{14} s_{13})\,,  \\
 \label{eq:A44} A_{44}
     & = c_{24}c_{34} c_{14}\,, 
 \end{align}
% ..........................................................................
independent of the mixing angle $\theta_{23}$, due to the leftmost positioning
of the matrix $R_{23}$ in the definition of the mixing matrix  in Eq.~(\ref{eq:U}).
The relations in Eq.~(\ref{eq:Us1}-\ref{eq:A44}) also evidence the following further features of our parameterization: 
IV)  For small values of  all the mixing angles involving the fourth mass eigenstate 
[$s^2_{i4 }\ll1\, (i=1,2,3)]$, the sterile flavor content 
is mostly distributed on  the fourth mass eigenstate ($U_{s4} \simeq 1$); 
V) For $\theta_{13} \simeq \theta_{14} \simeq 0$, the sterile content of the first two mass 
eigenstates essentially depends only on $\theta_{24}$, being in this limit $U_{s1}^2 + U_{s2}^2 \simeq A_{42}^2 = s_{24}^2$; 
VI) For  $\theta_{13} \simeq \theta_{14} \simeq \theta_{24} \simeq 0$, the mixing angle $\theta_{34}$ basically 
exchanges the amplitude of $U_{s3}$ with that of $U_{s4}$.%  

It should be stressed that our parameterization --- ``tailored'' for the solar sector ---
is different from that commonly adopted for the 4-flavor analyses of the atmospheric and long
baseline  neutrino 
data~\cite{Maltoni:2007zf, Adamson:2010wi}. In such a case, the mixing matrix is still
taken of the form  $U = BR_{14} R_{13} R_{12}$, but with a different choice of the ordering 
of the rotations entering $ B$ ($B_{atm} \equiv R_{34}R_{24} R_{23}$), which is dictated by the fact that 
$\nu_\mu \to \nu_\tau$ transitions are sensitive to the matrix elements
connecting $\nu_\mu$ and $\nu_\tau$  to the  new mass eigenstate $\nu_4$. 
Indeed, with such a choice, in the limit of small admixtures with the fourth state,
one has the approximate  expressions  $s_{24}^2 \simeq U_{\mu 4}^2$ and  $s_{34}^2 \simeq U_{\tau 4}^2$. 
In this regard, we observe that the parameterization we have adopted could be classified as of a ``mixed form'',
in between the two ones recognized in~\cite{Cirelli:2004cz} as more apt for studying those datasets 
possessing, respectively, a prevailing sensitivity to admixtures (of the new state $\nu_s \sim \nu_4$)
with the flavor eigenstates ($U_{\alpha 4} \ne 0$)  or with the mass ones ($U_{si} \ne 0$). 
Such a  particular form ensues from the peculiar properties of the solar sector
data, which are sensitive both to $U_{e4}$ (``favor-type'' admixture) and to the $U_{si}$'s (``mass-type'' admixtures).

\subsection{Four-flavor evolution}

The evolution  of the neutrino flavor eigenstates is governed by the Schr\"{o}dinger-like  equation
%........................................................................
\begin{equation}
\label{eq:4nuevol}
 i\, \frac{d}{dx}\left(\begin{array}{c}\nu_e\\ \nu_\mu \\ \nu_\tau \\ \nu_s \end{array}\right) = H_f
 \left(\begin{array}{c}\nu_e\\ \nu_\mu \\ \nu_\tau \\ \nu_s \end{array}\right)\,,
\end{equation}
%........................................................................
where the Hamiltonian
%........................................................................
\begin{equation}
H_f = H_f^{kin} + H_f^{dyn} = UKU^T + V(x)\,,
\label{eq:Hf}
\end{equation}
%........................................................................
has been split in the sum of a kinematical and a  dynamical term.  
In Eq.~(\ref{eq:Hf}) $K$  denotes the diagonal matrix containing the wavenumbers  
$k_i = m^2_i /2E \, (i=1,2,3,4)$ ($m_i$ and $E$ being the neutrino mass-squared  and 
energy respectively), while the matrix $V(x)$ incorporates the matter MSW 
potential~\cite{Wolfenstein:1977ue, smirnov}.
Barring irrelevant factors proportional to the identity, we can define the
diagonal matrix containing the three relevant wavenumbers as
%...........................................................................
\begin{align}
\label{eq:Vabstd}
  K  &  =  \mathrm{diag} (0, k_{sol}, k_{atm}, k_{new})\\
       &  \equiv \mathrm{diag}\Bigg( 0,\,\frac{\Delta m^2_{21}}{2E}, \, \frac{\Delta m^2_{31}}{2E}, \, \frac{\Delta m^2_{41}}{2E}\Bigg)\,,
 \end{align}
%...........................................................................
and the matrix encoding the matter effects, as  
%...........................................................................
\begin{equation}
\label{eq:V_matrix}
  V=\mathrm{diag}(V_{CC},\,0,\,0,\, -V_{NC})\,,
 \end{equation}
%...........................................................................
where
%...........................................................................
\begin{eqnarray}
V_{CC}
&=&
\sqrt{2} \, G_F \, N_e(x)\,,
\label{VCC}
\end{eqnarray}
%...........................................................................
is the charged-current interaction potential of the electron neutrinos with 
the background electrons having number density $N_e$, and
%.........................................................................
\begin{eqnarray}
V_{NC}
&=&
- \frac{1}{2} \sqrt{2} G_F N_n(x)\,,
\label{VNC}
\end{eqnarray}
%...........................................................................
is the neutral-current interaction potential (common
to all the active neutrino species) with the background 
neutrons having number density $N_n$. For later convenience, we  
also introduce the position-dependent parameter
 $r_x \equiv r(x)$  defined as the positive-definite ratio
%...............................................................................................
\begin{equation}
r_x = - \frac{V_{NC}(x)} {V_{CC}(x)}  = \frac{1}{2} \frac {N_n(x)}{N_e(x)}\,.
\end{equation}
%...............................................................................................

\subsection{Oscillation probabilities in vacuum}

In the case of propagation in vacuum, Eq.~(\ref{eq:4nuevol}) leads to the survival probability 
of electron (anti-)neutrinos  (which is relevant for the reactor experiment KamLAND%
%%%%%%%%%%%%%%%%%%%%%%%%%%%%%%%%%%%%%%%%%%%%%%%%%%%%%
\footnote{In this discussion we neglect the small matter effects induced by the interaction of the
electron antineutrinos with the Earth crust. However, for the sake of precision, we include these
effects in the numerical analysis presented in the next section.})
%%%%%%%%%%%%%%%%%%%%%%%%%%%%%%%%%%%%%%%%%%%%%%%%%%%%%
% ---------------------------------------------------------------------------------------------
\begin{equation}
P_{ee} = 1 - 4 \sum_{j>k} U_{ej}^2 U_{ek}^2 \sin^2 {\phi_{jk}} \,,
\end{equation}
% ---------------------------------------------------------------------------------------------
with the elements $U_{ei} \, (i =1,..., 4)$ determining the amplitudes of the oscillating terms
having phases  $\phi_{jk} = \Delta m^2_{jk}L/4E$ developed 
over the baseline $L$. In the hierarchical limit  $ k_{sol} \ll k_{atm}  \ll k_{new}$  the fast oscillations induced
by the two larger wavenumbers are completely averaged and the survival probability can be written as
% ---------------------------------------------------------------------------------------------
\begin{equation}
\label{Pee_vac_4nu}
P_{ee} = (1 - \sum_{k=3}^{4} U_{ek}^2)^2 P_{ee} ^{2\nu} +  \sum_{k=3}^{4} U_{ek}^4\,, 
\end{equation}
% ---------------------------------------------------------------------------------------------
where 
% ---------------------------------------------------------------------------------------------
\begin{equation}
P_{ee}^{2\nu} =  1 -  4 s^2_{12}  c^2_{12} \sin^2{\phi_{12}}\,,
\end{equation}
% ---------------------------------------------------------------------------------------------
is the well known 2-flavor expression of the survival probability in vacuum.
 Eq.~(\ref{Pee_vac_4nu}) shows that the presence of 
the ``far'' eigenstates $\nu_3$ and $\nu_4$ is felt as a lack of unitarity of the
$(\nu_1,\nu_2)$ sector and that there is an exact degeneracy between $U_{e3}$ and $U_{e4}$.
In our  parameterization of the mixing matrix  Eq.~(\ref{Pee_vac_4nu}) reads
% ---------------------------------------------------------------------------------------------
\begin{equation}
P_{ee} =  c_{14}^4  c_{13}^4 P_{ee} ^{2\nu} + c^4_{14}s_{13}^4 + s_{14}^4\,,
\end{equation}
% ---------------------------------------------------------------------------------------------
which implies an approximate degeneracy between the two (small) mixing angles
$\theta_{13}$ and $\theta_{14}$.

\subsection{Transition probabilities in matter}

Matter effects play a central role in the conversion of solar neutrinos and should be
incorporated following the treatment presented in Appendix~\ref{sec:app_msw}. It must be observed
that, although the solar data are mainly  sensitive to the survival probability of  the electron neutrinos,
they also posses a sensitivity to the transition probability into sterile states ($P_{es}$) through the 
neutral current  (NC)  measurements performed by the SNO experiment and, to a lesser extent, 
through the elastic scattering (ES) interactions exploited by SuperKamiokande and Borexino.

As shown in Appendix~\ref{sec:app_msw}, for the small values of the two mixing angles $\theta_{13}$ and 
$\theta_{14}$ we are considering, the propagation of solar neutrinos is adiabatic,
and the transition probabilities only depend upon their production and detection points.
Neglecting Earth-induced matter effects%
%%%%%%%%%%%%%%%%%%%%%%%%%%%%%%%%%%%%%%%%%%%%%%%
\footnote{Here, for simplicity, we neglect Earth matter effects which, however, are properly 
included in the numerical analysis following the prescription described in Appendix~\ref{sec:app_earth}.} 
%%%%%%%%%%%%%%%%%%%%%%%%%%%%%%%%%%%%%%%%%%%%%%%
one has the general expressions
%.................................................................................................................
\begin{equation}
\label{eq:Pee_adia}
  P(\nu_e \to \nu_\alpha) = \sum_{i = 1}^4 U^2_{\alpha i} (U^{m}_{ei})^2 \,\,\,\,\,\,\, (\alpha = e,\mu,\tau, s)\,,
\end{equation}
%..................................................................................................................
where the $U_{ei}^m$'s denote the mixing matrix elements involving
the electron neutrino  in the production point. 
As shown in Appendix A, in our parameterization these 
elements can be  simply  obtained by replacing in Eqs.~({\ref{eq:Ue1}-\ref{eq:Ue4})
the mixing angle $\theta_{12}$ in vacuum with the corresponding one in matter
% .................................................................................................................
\begin{align}
    \label{eq:Ue1m}  U_{e1}^m 
     &= c_{14}  c_{13} c_{12}^m\,, \\\
\label{eq:Ue2m}  U_{e2}^m
        &=  c_{14}  c_{13} s_{12}^m\,,    \\
\label{eq:Ue3m}  U_{e3}^m 
        & =  U_{e3} = c_{14}  s_{13}\,,  \\
\label{eq:Ue4m}  U_{e4}^m
        &= U_{e4} = s_{14} \,,
\end{align}
% .................................................................................................................
where $\theta_{12}^m$ can be calculated using the prescriptions provided in Eqs.~(\ref{eq:sin_12_matt}-\ref{eq:k_matt}).
In the numerical analysis we are going to present in this work, we will limit ourselves
to the simple case $\theta_{24} = \theta_{34} =0$. In this case, the matrix
A in Eq.~(\ref{eq:U}) takes the simpler form $A = R_{23} R_{14} R_{13}$, implying 
the following simplified expressions for the mixing elements involving the sterile flavor
% ....................................................................................................................
\begin{align}
    \label{eq:Us1_s} 
    U_{s1} & =  -s_{14} c_{13} c_{12}\,, \\
\label{eq:Us2_s}  
    U_{s2} & =   -s_{14} c_{13} s_{12}\,,  \\
\label{eq:Us3_s}
     U_{s3} &= -s_{14} s_{13}\,, \\
\label{eq:Us4_s} 
     U_{s4}  &  = c_{14}\,,
\end{align}
% .................................................................................................................
which show that, in this case, the small departure from unity of the sterile flavor content of the fourth 
mass eigenstate is redistributed (by unitarity) essentially to the first two mass eigenstates, 
leaving $U_{s3}^2 \sim 0$.  
Using Eqs.~(\ref{eq:sin_12_matt}-\ref{eq:k_matt}), one arrives at  the expressions
%...........................................................................
\begin{align}
\label{eq:Pee_4nu_13_14}
P_{ee} & =  c^4_{14}c^4_{13} \bar P_{ee}^{2\nu}  + c^4_{14} s^4_{13} + s^4_{14}\,, \\
\label{eq:Pes_4nu_13_14}
P_{es}  & =   c^2_{14}  s^2_{14} c^4_{13} \bar P_{ee}^{2\nu} + c^2_{14} s^2_{14}s^4_{13}  +  c^2_{14}s^2_{14}  \,, 
\end{align}
%...........................................................................
where 
%...........................................................................
\begin{align}
\label{eq:Pee_2nu_rot}
\bar P_{ee}^{2\nu} & =  c^2_{12} (c_{12}^m)^2  + s^2_{12}  (s_{12}^m)^2\,, 
\end{align}
%...........................................................................
is the well known expression of the 2-flavor survival probability in the adiabatic regime~\cite{Kuo:1989qe}
but with the mixing angle $\theta_{12}^m$ obtained using the rescaled charged-current potential 
$V_{CC} \to (c^2_{14}  + r_x s^2_{14}) c^2_{13} V_{CC}$ (see Appendix A). In the case of
 $\theta_{14} =0$, Eqs.~(\ref{eq:Pee_4nu_13_14}-\ref{eq:Pes_4nu_13_14})  
return the standard 3-flavor expressions~\cite{Shi:1991zw,Fogli:1993ck}
%...........................................................................
\begin{align}
\label{eq:Pee_4nu_13}
P_{ee}^{3\nu} & =  c^4_{13} \bar P_{ee}^{2\nu} + s^4_{13}\,, \\
P_{es}^{3\nu}  & =   0\,,
\end{align}
%...........................................................................
with the well known rescaling of the standard MSW potential $V_{CC} \to c^2_{13}V_{CC}$~\cite{Shi:1991zw,Fogli:1993ck}.
Conversely, in the case  ($\theta_{13} = 0$, $\theta_{14} \ne 0$), one gets 
%...........................................................................
\begin{align}
\label{eq:Pee_4nu_14}
P_{ee}(\theta_{13} = 0) & =  c^4_{14} \bar P_{ee}^{2\nu} + s^4_{14} \,,\\
\label{eq:Pes_4nu_14}
P_{es}(\theta_{13} = 0)   & =   c^2_{14} s^2_{14} (\bar P_{ee}^{2\nu} + 1)\,,
\end{align}
%...........................................................................
with the rescaled potential $V_{CC} \to (c^2_{14}  + r_x s^2_{14}) V_{CC}$.
We observe that, while the form of the electron neutrino survival probability is identical
to that obtained in the 3-flavor case, modulo the replacement $\theta_{13} \to \theta_{14}$ and
a small change in the rescaling factor of the potential, in this case, there is a non-zero 
transition probability into sterile states. In all cases, the change in the potential is very
small and, analogously to the 3-flavor limit~\cite{Fogli:2005cq}, it introduces only a mild 
(undetectable) energy dependence of the (significant) energy-independent suppression 
induced by the kinematical factor $c^4_{14} c^4_{13} \sim 1 - 2 s^2_{14} -2  s^2_{13}$ 
appearing in  Eq.~(\ref{eq:Pee_4nu_13_14}). 

%%%%%%%%%%%%%%%%%%%%%%%%%%%%%%%%%%%%%%%%%%%
\begin{figure}[t!]
\vspace*{-6.5cm}
\hspace*{1.4cm}
\includegraphics[width=21.0 cm]{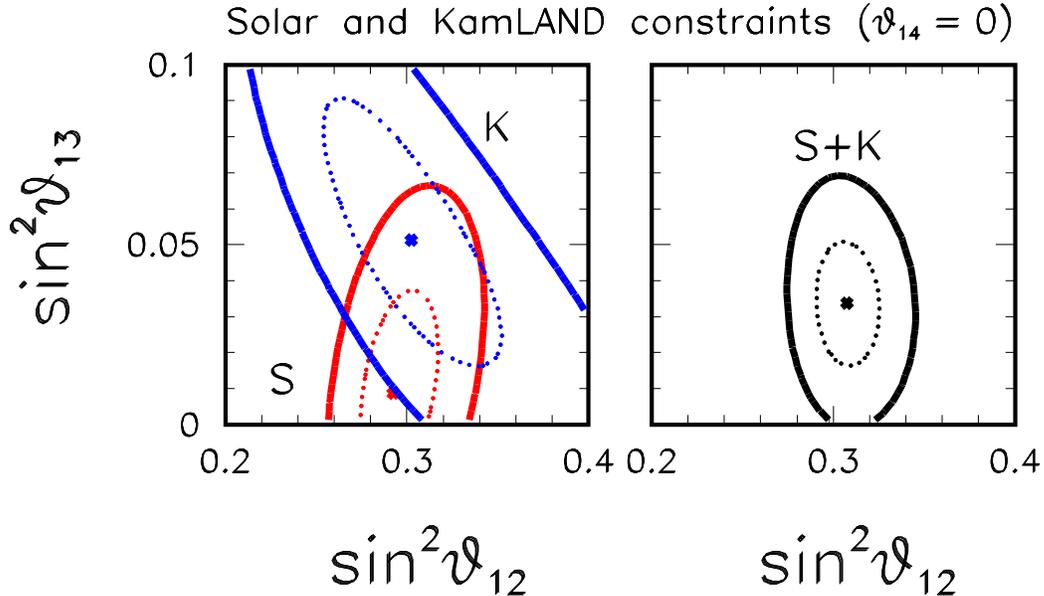}
\vspace*{-6.0cm}
\caption{Region allowed in the [$\sin^2\theta_{12}, \sin^2\theta_{13}$]  plane for $\theta_{14} =0$, after marginalization
of $\Delta m^2_{12}$ as constrained by KamLAND,  separately  (left panel) by solar (S) and KamLAND (K) 
data and by their combination (right panel).  In both panels  it has  been set $\theta_{24} = \theta_{34} =0$.
The contours refer to $\Delta \chi^2 =1$ (dotted line) and $\Delta \chi^2 = 4$ (solid line).
\label{fig1}}
\end{figure}  
%%%%%%%%%%%%%%%%%%%%%%%%%%%%%%%%%%%%%%%%%%%%%

\section{Numerical results}

In our analysis we have included the data from the 
Chlorine (Cl) experiment~\cite{cleveland:1998nv}, the
Gallium (Ga) detectors SAGE~\cite{abdurashitov:2002nt} and
GALLEX/GNO~\cite{Hampel:1998xg,Altmann:2005ix,kirsten2008retrospect},
Super-Kamiokande~\cite{fukuda} (SK),
all the three phases of the Sudbury Neutrino Observatory~\cite{ahmad,Ahmed:2003kj,Aharmim:2005gt,Aharmim:2008kc}
(SNO), and Borexino~\cite{Collaboration:2011rx,Bellini:2008mr} (BX). We have not included 
the spectral information provided by the SNO Low Energy Threshold Analysis~\cite{Aharmim:2009gd},
since this can be used only under the assumption of unitary
conversion among the active neutrino species ($P_{ee} + P_{e\mu} + P_{e\tau} = 1$),
which is violated in the presence of transitions into new sterile states. 
For the sake of precision, we have also incorporated the small
Earth-induced matter effects, following the treatment presented in Appendix~{\ref{sec:app_earth}.
Concerning KamLAND,  we have included  in our analysis the latest data released in~\cite{Gando:2010aa}.
For definiteness, we have adopted  the new improved reactor flux determinations~\cite{Mueller:2011nm}.
All plots will refer to this case and, when appropriate, we will comment in the
 text on the differences  that would arise with a different choice of the reactor fluxes. 
In all numerical computations we have set  $\theta_{24} = \theta_{34} =0$, commenting 
at the end of the Section on the possible role of these parameters. Therefore, 
the parameter space spanned by our analysis will involve the solar mass-spliting
$\Delta m^2_{12}$ and the three mixing angles ($\theta_{12}, \theta_{13}, \theta_{14}$).

We start our numerical study considering the familiar three-flavor case
($\theta_{13} \ne 0, \theta_{14} =0$), in which the results of the analysis depend on the
three parameters ($\Delta m^2_{12}, \theta_{12}, \theta_{13}$). This case
will serve as a useful term of comparison for the more general results  of the $4\nu$ analysis.  
In the left panel of Fig.~1 we show the region allowed by solar (S) and KamLAND (K) 
in the plane spanned by the two mixing angles, having marginalized away
the solar mass splitting in the region determined by KamLAND.
Respect to previous analyses~\cite{Fogli:2008jx,Balantekin:2008zm,Schwetz:2008er, Ge:2008sj, GonzalezGarcia:2010er, Gando:2010aa},
the KamLAND data  {\em taken alone} now tend to prefer  values of $\theta_{13}>0$
(see also~\cite{Mention:2011rk,Schwetz:2011qt}). This behavior
can be traced to our adoption of the new (higher) reactor fluxes~\cite{Mueller:2011nm}.
In fact,  according to Eq.~(\ref{eq:Pee_4nu_13}), a larger value of $\theta_{13}$ is now required
to suppress the bigger total rate induced by the new higher fluxes.
Furthermore, similarly to previous analyses~\cite{Fogli:2008jx, Balantekin:2008zm,Schwetz:2008er, Ge:2008sj, GonzalezGarcia:2010er,Gando:2010aa}, for $\theta_{13}>0$ the values of the mixing angle
$\theta_{12}$ identified by  the solar and KamLAND  experiments
are in better agreement due to the opposite-leaning correlations
exhibited  by their respective contours, giving rise to an enhanced preference
for non-zero $\theta_{13}$ in their combination (right panel). 
We find that the 2-flavor case ($\theta_{13} =0$) is disfavored at the 1.8~$\sigma$ level
(which is reduced to 1.3~$\sigma$ using the old reactor fluxes).

%%%%%%%%%%%%%%%%%%%%%%%%%%%%%%%%%%%%%%%%%%%
\begin{figure}[t!]
\vspace*{-6.5cm}
\hspace*{1.6cm}
\includegraphics[width=21.0 cm]{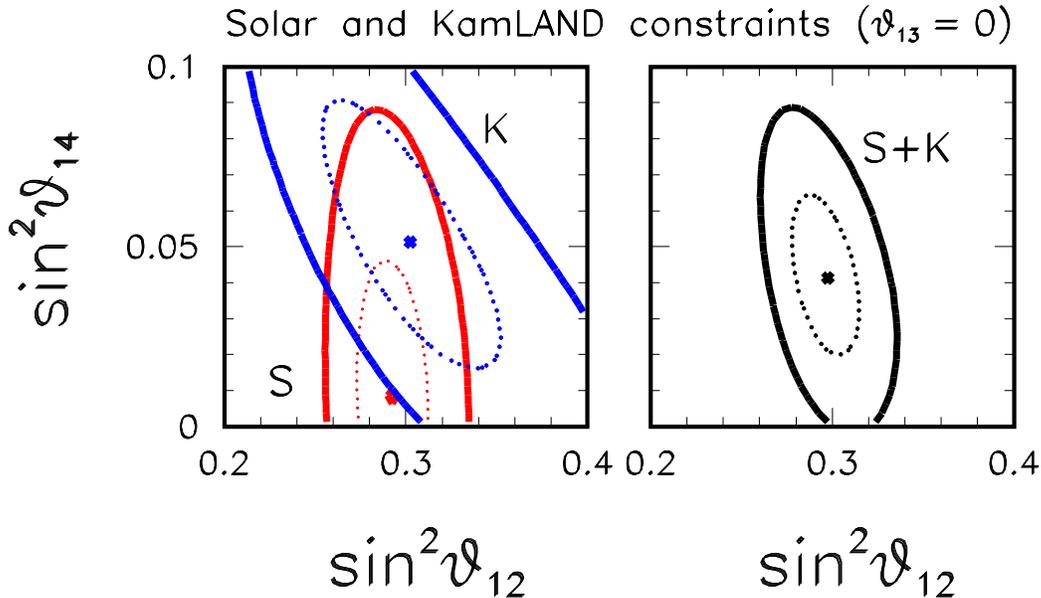}
\vspace*{-6.0cm}
\caption{Region allowed in the [$\sin^2\theta_{12}, \sin^2\theta_{14}$] plane for $\theta_{13} =0$, after marginalization
of $\Delta m^2_{12}$ as constrained by KamLAND,  separately  (left panel) by solar (S) and KamLAND (K) 
data and by their combination (right panel).  In both panels  it has  been set $\theta_{24} = \theta_{34} =0$.
The contours refer to $\Delta \chi^2 =1$ (dotted line) and $\Delta \chi^2 = 4$ (solid line).
\label{fig2}}
\end{figure}  
%%%%%%%%%%%%%%%%%%%%%%%%%%%%%%%%%%%%%%%%%%%%%

As as second step we switch on only the mixing angle $\theta_{14}$,
setting $\theta_{13} = 0$. In this case, the results of the analysis depend on the three parameters
($\Delta m^2_{12}, \theta_{12}, \theta_{14}$), whose allowed regions
are displayed in Fig.~2. As discussed in Sec.~II,  KamLAND cannot
distinguish $\theta_{13}$ from $\theta_{14}$ and, as a result, the region
identified by such an experiment is identical to that found  in the 3-flavor case.
In contrast, the region determined by the solar data is slightly different from the corresponding
one identified in the 3-flavor case. In particular, we see that the correlation in
the [$s^2_{12}, s^2_{14}]$  plane is different from that exhibited
in the [$s^2_{12}, s^2_{13}$]  plane.  To understand this point,  
it is useful to write the relations connecting the total
fluxes observed in the solar neutrino experiments to the electron neutrino
survival probability and their transition probability into sterile states.
Taking into account that $P_{ee} + P_{e\mu} + P_{e\tau} + P_{es} = 1$, we have  
%..........................................................................
\begin{align}
\label{eq:CC}
{\phi^ \mathrm{CC}}  & =  \Phi_B \langle P_{ee} \rangle\,, \\	
\label{eq:ES}
{\phi^\mathrm{ES}}   & =  \Phi_B (\langle P_{ee} \rangle  + r_\sigma (1- \langle P_{ee} \rangle -\langle P_{es} \rangle)\,,\\
\label{eq:NC}
{\phi^\mathrm{NC}}  & =  \Phi_B  (1 - \langle P_{es} \rangle)\,,
\end{align}
%..........................................................................
where $\Phi_B$  is the solar $^8$$B$ neutrino flux expected in the
absence of oscillations,
while $\phi^\mathrm{X}$  (X = CC, ES, NC) is the flux expected in
charged-current (CC),  elastic-scattering (ES) and neutral-current (NC)
reactions. The symbol $\langle \rangle$ denotes the average over  appropriate response
functions~\cite{Villante:1998pe, Fogli:2001nn}, which depend on the experiment and on the 
specific reaction considered. We remind that:  I) The radiochemical
experiments (Cl, Ga) employ a CC absorption reaction and are thus sensitive
only to $P_{ee}$; II) The BX and SK detectors make use of an ES reaction
and can probe also $P_{es}$, although with sensitivity
reduced by the small ratio $r_\sigma   \sim 1/6$
of the energy-averaged ES cross-sections of $\nu_{\mu,\tau}$ and $\nu_e$;
III) The SNO experiment detects the solar neutrinos using all the three reactions (CC, ES, NC), 
thus possessing a pronounced sensitivity to $P_{es}$  through the NC process. 

In the 3-flavor case, the sensitivity of the solar data {\em taken alone} to the mixing angle  $\theta_{13}$ 
 arises from an interplay of low-energy (LE) and high-energy (HE) (essentially the SNO CC/NC ratio)
data~\cite{Goswami:2004cn,Fogli:2006fu}, engendered by a different dependence of  the survival probability 
on the two mixing angles  $\theta_{12}$ and $\theta_{13}$ in the two regimes.
From Eqs.~(\ref{eq:CC}-\ref{eq:NC}), taking into account that in the LE (vacuum-like) 
regime $\bar P_{ee}^{2\nu} \sim 1 - 2s^2_{12}c^2_{12}$ and in the HE (matter-dominated) 
regime $\bar P_{ee} ^{2\nu}\sim s^2_{12}$, and using Eq.~(\ref{eq:Pee_4nu_13}),
we have the following relations for the LE and HE fluxes 
%..........................................................................
\begin{eqnarray}
\label{eq:rate_LE_13}
\phi_\mathrm{LE}^\mathrm {CC} 		\propto& P_{ee}^\mathrm{LE}  \simeq  (1 - 2s^2_{13})  (1- 2 s^2_{12} c^2_{12})\,, \\
\label{eq:rate_HE_13}
\Bigg[\frac{\phi^\mathrm {CC}}{\phi^\mathrm{NC}}\Bigg]_{\mathrm {HE}}  \simeq & \hspace{-1.45cm} P_{ee}^\mathrm{HE}   \simeq (1 - 2 s^2_{13}) s^2_{12} \,.
\end{eqnarray}
%..........................................................................
The different relative sign of the two factors proportional to $s^2_{13}$ and $s^2_{12}$
in Eqs.~(\ref{eq:rate_LE_13}-\ref{eq:rate_HE_13}) gives rise to an opposite correlation
among these two parameters,  providing an enhanced sensitivity to small departures from zero
of the mixing angle $\theta_{13}$.
Since the SNO CC/NC ratio is measured with better precision than the LE flux
(essentially provided by the Ga experiments), the negative relative sign in Eq.~(\ref{eq:rate_HE_13})
 prevails in the global fit, giving rise to the positive overall correlation among the two mixing angles
appearing in the left panel of Fig.~1 in the curve designed with label ``S''.  Indeed, according to Eq.~(\ref{eq:rate_HE_13}), a bigger value of
$\theta_{13}$ is needed to counterbalance the effect of a larger $\theta_{12}$, in order
to keep  the CC/NC ratio at the fixed value determined by the SNO experiment.

In the case of $\theta_{14} \ne 0$, while we have an identical expression for
the expected flux in the LE limit (modulo the replacement $\theta_{13}\to \theta_{14}$),
the SNO  CC/NC ratio depends also on $P_{es}$. For small values of $\theta_{14}$,
using Eqs~(\ref{eq:Pee_4nu_14}-\ref{eq:Pes_4nu_14}), we have
%..........................................................................
\begin{eqnarray}
\phi^{\mathrm CC}_\mathrm {LE} 	        \propto& \hspace{-0.45cm} P_{ee}^\mathrm{LE}  \simeq  (1 - 2s^2_{14})  (1- 2 s^2_{12} c^2_{12})\,, \\
\Bigg[\frac{\phi^\mathrm {CC}}{\phi^\mathrm{NC}}\Bigg] _{\mathrm {HE}} \simeq& P_{ee}^\mathrm{HE} (1 + P_{es}^\mathrm{HE})  \simeq (1 - c^2_{12} s^2_{14}) s^2_{12} \,.
\end{eqnarray}
%..........................................................................
In this case the coefficient in front of the product $s_{14}^2 s_{12}^2$ in the 
SNO CC/NC ratio is still negative but three times smaller ($c^2_{12} \sim 0.7$)  
than the corresponding one appearing in the 3-flavor expression. 
As a result, in such a case,  the synergetic effect of the 
combination LE and HE data slightly decreases, with a 
lower sensitivity to $\theta_{14}$ and a weak negative overall correlation
in the [$s^2_{12}, s^2_{14}]$ plane (see the curve designed with label ``S''  in the left panel of Fig.~2),
as now the LE data can compete with the HE ones
in determining the relative sign in front of the  product $s_{14}^2 s_{12}^2$.
Similarly to the 3-flavor case, the values of the mixing angle $\theta_{12}$ identified, 
respectively, by solar and KamLAND  are in better agreement for $\theta_{14}>0$,
with an enhanced preference for non-zero values of this parameter in their combination (right panel
in Fig.~2). 
Also in this case we find that $\theta_{14} =0$ is disfavored at the 1.8~$\sigma$ level
(which is reduced to 1.3~$\sigma$ using the old reactor fluxes).

The following small differences appear between the two cases:
I) A weaker upper bound on  $\theta_{14}$ ($s^2_{14} < 0.089$ at the $2\sigma$ level)
with respect to that obtained for $\theta_{13}$ ($s^2_{13} < 0.070$ at the $2\sigma$ level);
II) A slightly bigger best fit value for $\theta_{14}$ ($s^2_{14} = 0.041$)
with respect  to that obtained for $\theta_{13}$ ($s^2_{13} = 0.033$).
 It is interesting to note that the best fit value obtained for $\theta_{14}$ practically coincides with that 
indicated by the VSBL reactor and Gallium  calibration anomalies taken in combination~\cite{Mention:2011rk}.
Therefore, combining the solar sector results with such data would reinforce
their preference for non-zero $\theta_{14}$, providing an overall indication, which
we roughly estimate to be around the $\sim 4\sigma$ level.

%%%%%%%%%%%%%%%%%%%%%%%%%%%%%%%%%%%%%%%%%%%
\begin{figure}[t!]
\vspace*{-1.5cm}
\hspace*{2.9cm}
\includegraphics[width=16.0 cm]{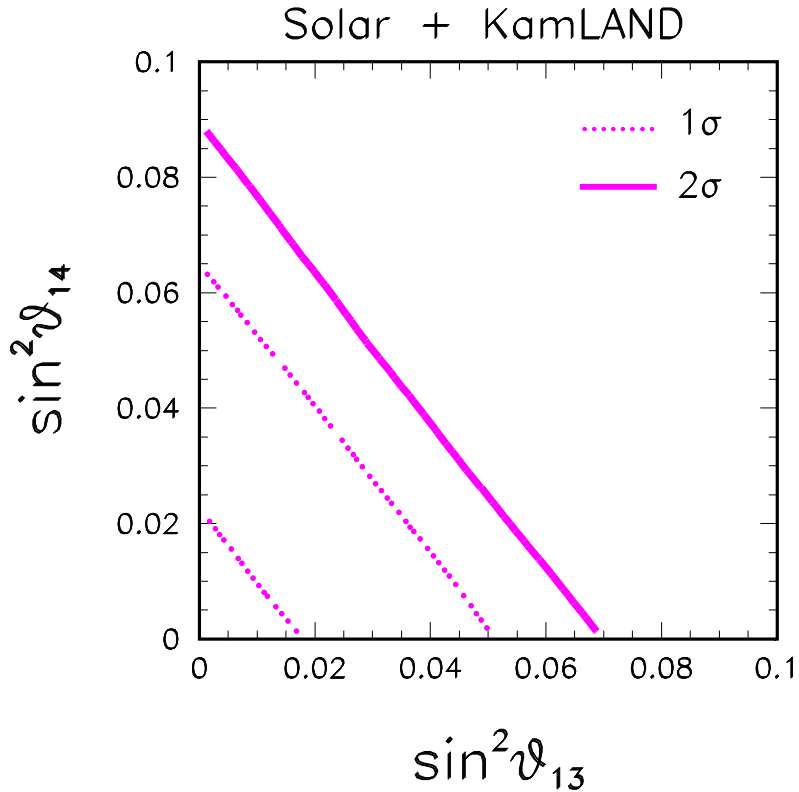}
\vspace*{-3.3cm}
\caption{Region allowed, after marginalization of $\Delta m^2_{12}$  and $\theta_{12}$,
by the combination of solar and KamLAND data (for $\theta_{24} = \theta_{34} =0 $).
 The contours refer to $\Delta \chi^2 =1$ (dotted  line) and
$\Delta \chi^2 = 4$ (solid line).
\label{fig2}}
\end{figure}  
%%%%%%%%%%%%%%%%%%%%%%%%%%%%%%%%%%%%%%%%%%%%%

As a third step of our numerical analysis, we have switched on both mixing angles
($\theta_{13} \ne 0, \theta_{14} \ne 0$). In Fig. 3 we show the region allowed 
by the combination of solar and KamLAND in the plane spanned by such two  
parameters, having marginalized away both the mass splitting $\Delta m^2_{12}$ 
and the mixing angle $\theta_{12}$. From this plot we see that there is a 
complete degeneracy among the two parameters. In practice, this dataset
is basically sensitive to the combination $U_{e3}^2 + U_{e4}^2$, 
the small deviations  from this behavior being induced by the SNO NC measurement.
Therefore, the solar sector data, while indicating a weak preference for non-zero
mixing with the ``far'' eigenstates $\nu_3$ and $\nu_4$, cannot distinguish
between them. The same conclusion holds true if additional sterile neutrinos
are considered, as shown in  Appendix B.

We conclude this section with a final remark. As in our analysis we have set $\theta_{24} = \theta_{34} =0$,
the question arises as to whether a  non-zero value of such parameters can alter the basic conclusions
of the analysis. With this purpose, we have performed the analysis considering this possibility and we
have found that, although the correlations observed in Figs.~(1,2) can slightly change, the degeneracy 
among the two parameters $\theta_{13}$ and $\theta_{14}$ persists, as one may have expected, 
since the additional freedom given to the system can only worsen its preexisting degeneracies.
We postpone to a future work the investigation of the constraints attainable on the mixing angles 
$\theta_{24}$ and $\theta_{34}$, noticing that in this case one should take into account the
(small) effects of the CP violating phases entering the lepton mixing matrix, as discussed in  Appendix D.

\section{Conclusions}

Motivated by the recent experimental findings which point towards the existence 
of new light sterile neutrino species, we have explored their impact on the solar sector 
phenomenology.  Working in a CPT-conserving 3+1 scheme, we have considered
the perturbations induced by a non-negligible mixing of the electron neutrino 
with a fourth sterile neutrino specie. Our quantitative analysis shows that the present  
data posses a sensitivity to the amplitude of the lepton mixing matrix element $U_{e4}$, 
which is comparable to that achieved  on the standard matrix element $U_{e3}$.
In addition, our analysis evidences that, in a 4-flavor framework, the current preference for $|U_{e3}|\ne 0$ is indistinguishable from that for $|U_{e4}|\ne0$, having both a similar statistical significance 
(which is $\sim 1.3 \sigma$  adopting the old reactor fluxes determinations, and $\sim 1.8\sigma$ 
using their new estimates.)  Such a degeneracy naturally extends to the more general $3+s$ schemes,
implying that in these frameworks,  the present hint for non-zero $\theta_{13}$ must be 
reinterpreted  as a preference for a non-zero mixing of the electron neutrino 
  with mass eigenstates distant from the ($\nu_{1},\nu_{2}$) ``doublet''. Different kinds of probes are needed
  in order to discriminate whether such an admixture --- if effectively confirmed to be non-zero --- is realized
   with the third standard mass eigenstate $\nu_3$ or with new neutrino specie(s).  

\section*{Acknowledgments}
We are grateful to  J.W.F. Valle for stimulating exchanges of ideas held at
the Instituto de F\'{\i}sica Corpuscular in Valencia,
where this work was initiated. We also thank E. Lisi, G.G. Raffelt and A.Yu. Smirnov
for precious discussions and A.M. Rotunno for useful information on the KamLAND
analysis. We thank  the organizers of the {\em 46th Rencontres de Moriond on Electroweak
Interactions and Unified Theories} held in La Thuile --- where 
preliminary results of this work were presented --- for kind hospitality.
Our work  is supported by the DFG Cluster of Excellence on the ``Origin and Structure of the Universe''.

\appendix

\section {The MSW effect in a 3+1 scheme}
\label{sec:app_msw}

To treat the MSW effect in a 3+1 scheme it is convenient to introduce the new basis
%..........................................................................
\begin{equation}
\bar \nu = A^T\nu
\,,
\label{eq:nu_bar}
\end{equation}
%..........................................................................
where
%..........................................................................
\begin{equation}
A = R_{23}  R_{24} R_{34} R_{14} R_{13} \equiv U R_{12}^T \,,
\end{equation}
%..........................................................................
is the same matrix defined in Eq.~(\ref{eq:U}).
In this new basis the Hamiltonian assumes the form
%.................................................................................................................
\begin{equation}
\label{eq:Hrot2}
  \bar H = \bar H^{kin} + \bar H^{dyn} = R_{12} K R_{12}^T + A^T  V A\,, 
\end{equation}
%..........................................................................
and, in the hierarchical limit $ k_{sol} \ll k_{atm}  \ll k_{new}$, similarly to
the 3-flavor case~\cite{Kuo:1989qe}, one can reduce the dynamics to that of an effective $2\nu$ system.
Indeed, from Eq.~(\ref{eq:Hrot2}) one has that the $(3,3)$ and $(4,4)$ entries of $\bar H$,
being proportional to $k_{atm}$ and $k_{new}$,  are much bigger than 
all the other ones, and at the zeroth order in the small quantities $V_{CC}/k_{atm}$,
$V_{NC}/k_{atm}$,  $V_{CC}/k_{new}$, $V_{NC}/k_{new}$, the third and fourth eigenvalues 
of $\bar H$ are much larger than the first two ones. As a result,  
the states $\bar \nu_3 = \nu_3$ and $\bar \nu_4 =  \nu_4$  evolve independent
one of each other and, more importantly, completely decoupled from  $\bar \nu_1$ and  $\bar \nu_2$. 
Extracting from $\bar H$ the submatrix with indices $(1,\,2)$ one obtains the $2\times2$ Hamiltonian 
%.................................................................................................................
\begin{equation}
\label{eq:H_rot}
  \bar H_{2\nu} = \bar H^{kin}_{2\nu} + \bar H^{dyn}_{2\nu}\,, 
\end{equation}
%..........................................................................
governing the  evolution of the ($\bar \nu_1,\bar \nu_2$) system,
whose dynamical part has the form 
%..........................................................................
\begin{equation}
\label{eq:H_rot_general}
    \bar H^{dyn}_{2\nu} = V_{CC}(x)
    \begin{pmatrix}
        A_{11}^2 +  r_x\, A_{41}^2\,  & A_{11}A_{12} + r_x A_{41} A_{42} \\
        A_{11}A_{12} + r_x A_{41} A_{42}\,   & A_{12}^2  +r_x A_{42}^2 \, 
    \end{pmatrix}
\,.   
\end{equation}
%..........................................................................
In our parameterization the rotation $R_{23}$ appears as the leftmost matrix in A,  
and therefore the product $A^T V A$ does not depend on $\theta_{23}$ since $R_{23}$
commutes with the matrix V of the potential given in Eq.~(\ref{eq:V_matrix}).
Furthermore, the matrix A contains the product $R_{14} R_{13}$ as its rightmost factor, 
implying  that its element $A_{12} $ is equal to zero. Therefore,  the $2\times2$ Hamiltonian in
Eq.~(\ref{eq:H_rot_general}) can be recast in the form
%.......................................................................................
%..........................................................................
\begin{equation}
\label{eq:H_dyn_rot_ab}
    \bar H^{\,dyn}_{2\nu} = V_{CC}(x)
    \begin{pmatrix}
        \gamma^2 +  r_x\, \alpha^2\,  & r_x\, \alpha\beta\,  \\
        r_x\, \alpha\beta\,         & r_x\, \beta^2\, 
    \end{pmatrix}
\,,    
\end{equation}
%..........................................................................
where the parameters ($\alpha, \beta, \gamma$) are defined as
%..........................................................................
\begin{align}
   \label{eq:alfa_app} \alpha
        &=  A_{41} = c_{24} (s_{34} s_{13} - c_{34} s_{14} c_{13})   \,, \\
 \label{eq:beta_app} \beta
    &= A_{42} =  - s_{24} \,,\\
\label{eq:gamma_app} \gamma
        &=  A_{11} = c_{13}  c_{14} \,,   
\end{align}
% ..........................................................................
which, according to Eqs.~(\ref{eq:Us1}-\ref{eq:Us4}), are related to the mixing matrix elements as follows
%..........................................................................
\begin{align}
    U_{s1}   & = \alpha \, c_{12} - \beta \, s_{12} \,,\\
    U_{s2}  & =  \alpha\, s_{12}  + \beta \, c_{12} \,, 
\end{align}
%..........................................................................
showing that the sum of  $\alpha^2$ and  $\beta^2$ 
represents  the sterile content of the ($\nu_1, \nu_2$) ``doublet''
%..............................................
\begin{align}
    \alpha^2 + \beta^2 = U_{s1}^2 + U_{s2}^2\,.
\end{align}
%.........................................................................
Furthermore, using Eqs.~(\ref{eq:Ue1_gen}-\ref{eq:Ue4}) we have
%..............................................
\begin{align}
    \gamma^2 = U_{e1}^2 + U_{e2}^2 = 1 - U_{e3}^2 - U_{e4}^2\,. 
\end{align}
%.........................................................................

It is instructive to consider the following limit cases:

\renewcommand{\theenumi}{\Roman{enumi}}

\begin{enumerate}

\item For vanishing mixing with the fourth neutrino specie ($\theta_{14} = \theta_{24} = \theta_{34} =  0$)
one has $\alpha = \beta = 0$ and $\gamma = c_{13}$,  thus recovering the standard 3-flavor result~\cite{Shi:1991zw,Fogli:1993ck}, in which the admixture with the third state $\nu_3$ induces  the simple rescaling of the potential $V_{CC} \to c^2_{13} V_{CC}$;

\item When only  the mixing angle $\theta_{14}$ is different from zero,
$\alpha = - s_{14}$,  $\beta = 0$ and  $\gamma = c_{14}$,   with
the position-dependent rescaling of the standard potential $V_{CC} \to (c^2_{14} +r_x s^2_{14}) V_{CC}$;

\item If both $\theta_{13}\ne0$ and $\theta_{14}\ne0$  (and $\theta_{24} = \theta_{34} =  0$),
$\alpha = -s_{14} c_{13}$, $\beta = 0$, $\gamma = c_{13}c_{14}$, with
the position-dependent rescaling of the standard potential $V_{CC} \to (c^2_{14} +r_x s^2_{14})c^2_{13} V_{CC}$;

\item In the case  of no admixture of the electron neutrino with the ``far'' states $\nu_3$ and $\nu_4$
($\theta_{13} = \theta_{14} = 0$), one has $\alpha = 0$,  $\beta = -s_{24}$ and $\gamma =1$,  with  
the position-dependent rescaling of the standard MSW potential $V_{CC} \to (1- r_x s_{24}^2) V_{CC}$, 
in agreement with the result found in~\cite{Dooling:1999sg} for this particular case.%. 
%%%%%%%%%%%%%%%%%%%%%%%%%%%%%%%%%%%%%%%%%%%%%%%%%
\footnote{In the different parameterization adopted in~\cite{Dooling:1999sg} for the lepton mixing matrix,
the role of $s_{24}^2$  is taken by $c^2_{23} c^2_{24}$. In both parameterizations, the
rescaling factor has (obviously) the same physical interpretation, being determined in both cases by
the sterile content of the ($\nu_1, \nu_2$) sector.}
%%%%%%%%%%%%%%%%%%%%%%%%%%%%%%%%%%%%%%%%%%%%%%%%%

\end{enumerate}

In the general case the modifications are less obvious than a simple rescaling of the standard
potential, as the $2\times2$ Hamiltonian in Eq.~(\ref{eq:H_dyn_rot_ab}) can contain both new 
diagonal and off-diagonal terms.
The matrix $\bar H_{2\nu}$ in Eq.~(\ref{eq:H_rot}) will be diagonalized by a $2\times2$ rotation 
%..........................................................................
\begin{equation}
    R_{12}^{m\, 2\times 2}(x)
    = 
\begin{pmatrix}
       \cos \theta_{12}^m  & \sin \theta_{12}^m \\
        -\sin \theta_{12}^m & \cos \theta_{12}^m 
    \end{pmatrix}\,,    
\end{equation}
%..........................................................................
which defines the mixing angle in matter $\theta_{12}^m$ 
that, in general, will depend on all the mixing angles
except for $\theta_{23}$. Therefore,  if we define the $4\times4$ rotation in the $(1,2)$ plane 
%%..........................................................................
\begin{equation}
    R_{12}^m (x)
    = 
\begin{pmatrix}
       R_{12}^{m\, 2\times 2}  & 0 \\
       0 & I^{2\times2}
    \end{pmatrix}\,,    
\end{equation}
%..........................................................................
the starting 4-dimensional Hamiltonian 
in Eq.~(\ref{eq:Hf}) will be diagonalized  by the matrix
%.................................................................................................................
\begin{equation}
\label{eq:Um}
U^m= A R_{12}^m\,,
\end{equation}
%..........................................................................
which connects the flavor eigenstates to the instantaneous energy eigenstates in matter
%..........................................................................
\begin{equation}
    \begin{pmatrix}
        \nu_e \\
        \nu_\mu\\
        \nu_\tau\\
        \nu_s         
    \end{pmatrix}
    =  U^m
\begin{pmatrix}
        \nu_1^m \\
        \nu_2^m \\
        \nu_3^m \\
        \nu_4^m
            \end{pmatrix}\,.  
\end{equation}
%..........................................................................
The mixing angle in matter $\theta_{12}^m$ is given by 
%..........................................................................
\begin{align}
   \label{eq:sin_12_matt} 
    \frac{k_m}{k} \sin  2\theta_{12}^m &=  \sin 2\theta_{12} + 2v_x r_x \alpha \beta \,, \\
    \label{eq:cos_12_matt} 
    \frac{k_m}{k} \cos  2\theta_{12}^m   &=  \cos 2\theta_{12} - v_x \gamma^2 - v_x r_x (\alpha^2 -\beta^2) \,,
 \end{align}
% ..........................................................................
where  $k\equiv k_{sol}$ and the neutrino wavenumber in matter $k_m$ is defined by
%..........................................................................
\begin{align}
   \label{eq:k_matt} 
     \frac{k_m^2}{k^2} =  [\cos 2\theta_{12} - v_x \gamma^2 - v_x r_x
     (\alpha^2 -\beta^2)]^2
      +   [\sin 2\theta_{12} + 2v_x r_x \alpha \beta] ^2\,,
 \end{align}
% ..........................................................................
with $v_x = V_{CC}(x)/k$. 
It should be noted that for $\alpha\beta <0$ {\em both} terms in Eq.~(\ref{eq:k_matt}) may become 
{\em simultaneously} small, even for large values of $\theta_{12}$. In particular, if at a point $x$ along the neutrino trajectory 
 the {\em two}  conditions%
 %%%%%%%%%%%%%%%%%%%%%%%%%%%%%%%%%%%%%%%%%%%%%
\footnote{Equations~(\ref{eq:vx_res}, \ref{eq:tan_12_res})  generalize  analogous conditions
introduced in~\cite{Bergmann:1997mr} in the context of solar neutrino conversion in the 
presence of non-standard neutrino interactions of the flavor-changing type. 
In that study, focused on the case of small mixing angles, it was shown that non-adiabatic effects are enhanced when: I) $v_x = \cos 2 \theta_{12} \simeq 1$ (corresponding to the common resonance condition); and
 II) $\tan 2 \theta_{12} = - 2\eps_{ea}$, where the coupling  constant $\eps_{ea}$, 
 parametrizing the strength of the new interaction between the two different flavors $\nu_{e}$ and  $\nu_{a}$, plays the role of our off-diagonal parameter $r_x \alpha \beta$. In the presence of additional flavor-diagonal 
 interaction terms of the type
 $\eps_{ee}$ and $\eps_{aa}$ one would have the more general conditions:
 I) $[1/v_x^2 = [(1+ \eps_{ee} - \eps_{aa})^2 + 4\eps_{ea}^2] $;  and 
  II) $\tan 2 \theta_{12} = - 2\eps_{ea}/ (1+ \eps_{ee} - \eps_{aa})$. In our case the diagonal
  terms correspond to $\eps_{ee} = -1 + \gamma^2 + r_x \alpha^2$ and $\eps_{aa} =   r_x\beta^2$
  [see Eq.~(\ref{eq:H_dyn_rot_ab})].}
  %%%%%%%%%%%%%%%%%%%%%%%%%%%%%%%%%%%%%%%%%%%%%%..........................................................................
\begin{align}
  \label{eq:vx_res} 
  \frac{1}{v_x^2} &= [\gamma^2 + r_x (\alpha^2 -\beta^2)] ^2 + 4 r_x^2 \alpha^2\beta^2 \,,\\
   \label{eq:tan_12_res} 
    \tan 2\theta_{12}&= \frac{2r_{x}|\alpha\beta|}{\gamma^2 +r_{x} (\alpha^2-\beta^2)}\,,
    \end{align}
% ..........................................................................
are {\em both} satisfied, the difference between the two energy eigenstates in matter approaches zero ($k_m \to 0$).
In such a case one expects important non-adiabatic effects
encoded by a non-zero swapping probability $P_c \equiv P(\nu_2^m \to \nu_1^m) $ between the two 
energy eigenstates in matter.  However, it turns out that the second condition [Eq.~(\ref{eq:tan_12_res})]
cannot be realized  for realistic values of the parameters involved in the conversion of solar neutrinos. 
On the one hand, the mixing angle $\theta_{12}$ is
constrained to have relatively big values by the KamLAND experiment (see the left panels of Figs.~1-2), which 
sets the robust lower limit $\tan 2\theta_{12} \gtrsim 1.0$ (at the $4 \sigma$ level),
 {\em independent} of any kind of matter-effects.  On the other hand,  the possible excursion 
 of the right term in  Eq.~(\ref{eq:tan_12_res}) is severely constrained by the non-solar neutrino
 oscillation phenomenology and by the properties of the Sun. Indeed we have that:
I) In the Sun the ratio $r_x$ never exceeds the maximum value (assumed at the Sun center)
$r_{max}\simeq 0.25$; II) The mixing angles ($\theta_{13}, \theta_{14}$) are bounded
by the reactor experiments ($s^2_{13} \lesssim 0.1$ and $s^2_{14} \lesssim 0.1$).
These circumstances ensure that the numerator in the right term of  Eq.~(\ref{eq:tan_12_res}) 
is always small, while keeping  its denominator close to one. Allowing for arbitrary values of the
other two mixing angles ($\theta_{24}, \theta_{34}$), we estimate that the ratio on the right side of
Eq.~(\ref{eq:tan_12_res})  never exceeds $0.2$. Therefore, non-adiabatic effects are completely 
irrelevant in the problem under study, as we have explicitly checked by a numerical scan of the
 relevant  parameter space. 
We stress that the same conclusion would not hold if the assumption of CPT
invariance were abandoned, as it ensures that the mixing angles probed by the reactor
antineutrinos are identical to those involved in the conversion of the 
solar neutrinos. 

Considering an electron neutrino produced in the Sun, the probability
to detect it on the Earth with flavor $\alpha$ will be
%.................................................................................................................
\begin{equation}
\label{eq:Pee_adia_phases}
  P(\nu_e \to \nu_\alpha) = \sum_{i = 1}^4 |U_{\alpha i} e^{\xi_i} U^{m}_{ei}|^2 \,\,\,\,\,\,\, (\alpha = e,\mu,\tau, s)\,,
\end{equation}
%..................................................................................................................
where the $U^{m}_{ei}$'s are the mixing matrix elements 
calculated  in the production point, while the
$\xi_i\sim \Delta m^2_{i1} L/2E $ are the phases acquired by the energy eigenstates
during their propagation from the Sun center to the Earth surface ($L$ $\simeq$ 1 a.u.).
The  information contained in the (large) phases gets lost by the spatial average over the 
neutrino production zone  and by the energy smearing~\cite{Lisi:2000su}, 
and Eq~(\ref{eq:Pee_adia_phases}) reduces to
%.................................................................................................................
\begin{equation}
\label{eq:Pee_adia_app}
  P(\nu_e \to \nu_\alpha) = \sum_{i = 1}^4 |U_{\alpha i}|^2 |U^{m}_{ei}|^2 \,\,\,\,\,\,\, (\alpha = e,\mu,\tau, s)\,,
\end{equation}
%..................................................................................................................
which coincides with the Equation~(\ref{eq:Pee_adia}) used in Sec.~II.
In summary, the calculation of the transition probability is reduced to the
following steps: I) Given the four mixing angles ($\theta_{13},\theta_{14},\theta_{24},\theta_{34}$),
as defined in the parameterization in Eq.~(\ref{eq:U}), calculate the coefficients ($\alpha, \beta, \gamma$)
making use of Eqs.~(\ref{eq:alfa_app}-\ref{eq:gamma_app}); II) Determine how the mixing angle
$\theta_{12}$ in vacuum gets modified in matter applying Eqs.~(\ref{eq:sin_12_matt}-\ref{eq:k_matt})
for the expression of $\theta_{12}^m$;
 III) Deduce the electron neutrino mixing elements  $U_{ei}^m$  in matter from Eq.~(\ref{eq:Um});
IV) Derive the transition probabilities $P(\nu_e \to \nu_\alpha)$ using Eq.~(\ref{eq:Pee_adia_app}).

\section {Generalization to a $3+s$ scheme}
\label{sec:app_3+s}

The generalization to more than one sterile specie is straightforward and is obtained by observing that
the $2\times2$  matrix $\bar H^{dyn}_{2\nu}$ entailing the non-trivial dynamics is always given by the
submatrix with indeces (1,2) of  the matrix
%........................................................................
\begin{equation}
\bar H^{dyn} = A^T VA  \,,
\label{hamiltonian2}
\end{equation}
%........................................................................
where now $A$, still defined as $A = U R_{12}^T$
like in the 3+1 scheme, has dimension $3+s$,
and the matrix V is given by $diag (V_{CC}, 0,0, -V_{NC}, -V_{NC}, ...)$.
If the matrix A is taken with the product $R_{1,3+s} ... R_{14} R_{13}$ 
of the matrices involving the first index as its rightmost factor
(ensuring $A_{12} =0$), the effective $2\times2$ Hamiltonian $\bar H^{dyn}_{2\nu}$ 
has  the same form of Eq.~(\ref{eq:H_dyn_rot_ab}) provided that the 
(combinations of the)  three coefficients $(\alpha, \beta, \gamma)$ appearing in
Eq.~(\ref{eq:H_dyn_rot_ab}) are replaced  as follows 
%..........................................................................
\begin{align}
       \alpha^2 \to & \sum_{i=1}^s \alpha_i^2 \,,\\  
       \beta^2 \to & \sum_{i=1}^s \beta_i^2\,,  \\       
       \alpha \beta \to & \sum_{i=1}^s \alpha_i \beta_i\,,    
\end{align}
%..........................................................................
with 
%..........................................................................
\begin{align}
     \alpha_i    = & A_{3+i, 1}\,\,\,\,\, (i=1, ... ,  s)\,, \\
     \,\,\, \beta_i  = & A_{3+i,2}\,\,\,\,\,  (i =1, ... ,s)\,,
      \end{align}
%..........................................................................
while
%..........................................................................
\begin{align}
    \gamma^2 \to 1 - U_{e3}^2 - \sum_{i = 4}^{3+s} U_{ei}^2\,.
\end{align}
%.........................................................................
Form this last formula it is evident that the role of the electron neutrino mixing with
additional sterile species is completely symmetrical to that played by  the fourth one.

\section {Inclusion of Earth matter  effects}
\label{sec:app_earth}

We briefly review the analytical results pertaining Earth matter effects, for the sake
of completeness and self-consistency of the paper. In general, Earth matter effects
intervening prior to the detection of the solar neutrinos, can be implemented by 
the following substitution in Eq.~(\ref{eq:Pee_adia})
% ---------------------------------------------------------------------------------------------------------
\begin{equation}
U^2_{\alpha i} \to P_{\alpha i}  \,\,\,\,\, (i=1,2,3,4;\, \alpha = e,\mu,\tau,s)\,,
\end{equation} 
% ---------------------------------------------------------------------------------------------------------
where $P_{\alpha i} \equiv P(\nu_i \to \nu_\alpha)$ are the conversion probabilities
in the Earth of the mass eigenstates into the flavor ones.
Although the $P_{\alpha i}$'s can be calculated numerically, it is possible
to simplify their evaluation by reducing the $4\nu$ dynamics to that of 
a $2\nu$ system, in analogy with the solar-matter induced effects discussed in Appendix A.
In fact, the $P_{\alpha i}$'s, can always be written as
%..........................................................................
\begin{eqnarray}
\label{eq:P_earth_ai}
P_{\alpha i}  = \left |(A W R_{12})_{\alpha i} \right|^2 \equiv \left |(A Z_{12})_{\alpha i} \right|^2\,,
\end{eqnarray}
%..........................................................................
where, from right to left: I) The matrix $R_{12}$ rotates the initial 
mass eigenstates into the auxiliary flavor basis $\bar \nu$ defined in Eq.~(\ref{eq:nu_bar}); 
II) The matrix $W$ contains the (complex)  transition amplitudes among the
eigenstates of such new basis, whose non trivial dynamics is confined
to the ($\bar \nu_1,\bar \nu_2$) sector; III) Finally, the matrix A 
rotates back the auxiliary basis $\bar \nu$ to the standard flavor basis.  The $4\times4$ matrix $W$
contains in its (1,2) subblock the non-trivial information, as follows
%..........................................................................
\begin{equation}
    Z_{12} = WR_{12} = \begin{pmatrix}
        \sqrt  {P} e^{-i \xi}        & \sqrt  {1- P} e^{-i \eta}  & 0 & 0\\
         -\sqrt  {1- P} e^{i \eta} &  \sqrt  {P} e^{i \xi}  & 0 & 0\\
        0&0&1&0\\
        0&0&0&1 
    \end{pmatrix}
\,,    
\end{equation}
%..........................................................................
where the 2-flavor transition probability $P \equiv \bar P^{2\nu}_{e1} = 1- \bar P^{2\nu}_{e2}$  
and the two phases ($\xi,\eta$) must be calculated numerically,
by implementing the MSW Hamiltonian of the form given in Eq.~(\ref{eq:H_dyn_rot_ab}), with 
number densities $N_e(x)$ and  $N_n(x)$  evaluated 
along the neutrino trajectory in the Earth interior. From Eq.~(\ref{eq:P_earth_ai}) one
obtains for the transition probabilities into electron neutrinos
% ---------------------------------------------------------------------------------------------------------
\begin{align}
\label{eq:Pe1_gen}
P_{e1} & =   |A_{11} \, Z_{11} + A_{12} \, Z_{21}|^2\,, \\
\label{eq:Pe2_gen}
P_{e2} & =   |A_{11} \, Z_{12}  + A_{12} \, Z_{22}|^2\,, \\
\label{eq:Pe3_gen}
P_{e3} & =   A_{13}^2 \equiv U^2_{e3}  \,, \\
\label{eq:Pe4_gen}
P_{e4} & =   A_{14}^2 \equiv U^2_{e4} \,,
\end{align} 
% ---------------------------------------------------------------------------------------------------------
which, taking into account the expressions of the elements of the first row of the matrix A
($A_{11}, A_{12}, A_{13}, A_{14}  = c_{14} c_{13},  0,  c_{14}s_{13}, s_{14}$),  become
 % ---------------------------------------------------------------------------------------------------------
\begin{align}
\label{eq:Pe1}
P_{e1} & =  c^2_{14}c^2_{13}  \bar P_{e1}^{2\nu}\,,\\
\label{eq:Pe2}
P_{e2} & =   c^2_{14}c^2_{13}  \bar P_{e2}^{2\nu}\,,\\
\label{eq:Pe3}
P_{e3} & = c_{14}^2 s_{13}^2\,,\\
\label{eq:Pe4}
P_{e4} & = s_{14}^2\,,
\end{align} 
% ---------------------------------------------------------------------------------------------------------
which constitute the generalization of the 3-flavor expressions  
used in the literature.
It should be noted  that in our parameterization it is $A_{12} = 0$, and we can express the
$4\nu$  transition probabilities in terms of the 2-flavor transition {\em probability} $\bar P_{1e}^{2\nu}$,
the knowledge of the phases ($\xi, \eta$) being unnecessary.
For the transition probabilities into sterile neutrinos, one obtains the analogous expressions
% ....................................... .......................................................................
\begin{align}
    \label{eq:Ps1_gen} P_{s1} 
     &= |A_{41} \, Z_{11} + A_{42} \, Z_{21}|^2 \,, \\
\label{eq:Ps2_gen} P_{s2}
        &=   |A_{41} \, Z_{12}  + A_{42} \, Z_{22}|^2\,,  \\
\label{eq:Ps3_gen} P_{s3} 
        & = A_{43}^2 \equiv U_{s3}^2\,, \\
\label{eq:Ps4_gen}  P_{s4}
        &= A_{44}^2 \equiv U_{s4}^2\,,
\end{align}
% .................................................................................................................
which, in the limit of no matter effects ($Z_{12} = R_{12}$),
return the expressions of $U_{si}^2$ given in Eqs.~(\ref{eq:Us1}-\ref{eq:Us4}).
Now, differently form the case of the electron neutrinos, both elements $A_{41}$ and $A_{42}$ 
can be different from zero [see Eqs.~(\ref{eq:A41}-\ref{eq:A44})], and one cannot express the
4-flavor transition probabilities $P_{s1}$ and $P_{s2}$  in terms of the 2-flavor transition
{\em probability} $\bar P_{1e}^{2\nu}$, needing the complete information on the (complex) 
2-flavor transition {\em amplitudes}, having phases ($\xi, \eta$).
We close this appendix by stressing that the expressions given in Eqs.~(\ref{eq:Pe1_gen},\ref{eq:Pe4_gen})
and Eqs.~(\ref{eq:Ps1_gen},\ref{eq:Ps4_gen}) are valid for any parameterization of the mixing matrix
of the form $U = AR_{12}$. The validity of Eqs.~(\ref{eq:Pe1},\ref{eq:Pe4}) is instead restricted to the case
in which the matrix  $A$  is of the form $A = BR_{14}R_{13}$.

\section{Potential sensitivity to the CP violating phases}
\label{sec:app_phases}

For definiteness, in the rest of the paper we have restricted ourselves
to the case of vanishing CP violating phases. Here, we briefly
comment on the potential sensitivity of solar neutrinos to them. 
In the 3-flavor framework  one can always eliminate the CP
violating phase $\delta$ appearing in the mixing matrix from all the observable quantities involved in the
solar neutrino transitions~\cite{Kuo:1987km,Minakata:1999ze},
since the two following conditions hold: I) All the relevant information 
on the flavor conversion is contained in the survival probability  $P_{ee}$ of the electron neutrinos, 
as one cannot distinguish  $\nu_\mu$ from $\nu_\tau$ at low energies 
and by unitarity it is $P_{e\mu} + P_{e\tau} = 1 -P_{ee}$; 
II) The phase $\delta$ can be eliminated from the expression of $P_{ee}$ since
it can be rotated away from the MSW dynamics due to the particular  form of the potential~\cite{Kuo:1987km,Minakata:1999ze} (see also~\cite{Yokomakura:2002av, Balantekin:2007es}).

The same conclusion is not  true in the 4-flavor case since
the two conditions above are no more valid, as we briefly show.
Without loss of generality we can assign one of the three phases to the
 $(2,3)$ sector since, as we have seen in Sec.~II  and Appendix A,  
the associated mixing angle $\theta_{23}$ can be eliminated from the description
of the solar neutrino conversion.
We can then attribute the remaining two phases to the (1,3)
and (1,4) sectors, by defining the complex mixing matrix as 
%..........................................................................
\begin{equation}
\label{eq:U_complez}
U =   \tilde R_{23}  R_{24} R_{34} \tilde R_{14} \tilde R_{13} R_{12} \,, 
 \end{equation} 
%..........................................................................
where $\tilde R_{ij}$ is a $4\times4$ complex rotation in the $(i,j)$ plane,
formed by replacing the real $2\times2$ submatrix  in Eq.~(\ref{eq:R_ij_2dim}) with the complex one
%..........................................................................
\begin{equation}
    \tilde R^{2\times2}_{ij} =
    \begin{pmatrix}
        \tilde c_{ij} & \tilde s_{ij}^*  \\
         -\tilde s_{ij}  & \tilde c_{ij}
    \end{pmatrix}
\,,    
\end{equation}
%..........................................................................
with $\tilde c_{ij} \equiv \cos \theta_{ij}$, $\tilde s_{ij} \equiv \sin \theta_{ij} e^{i\delta_{ij}}$.
The two phases $\delta_{13}, \delta_{14}$ will appear in the expression of the transition 
probability $P_{es}$ in Eq.~(\ref{eq:Pee_adia_phases})
through the (now complex) elements $U_{si}$ [see  Eqs.~(\ref{eq:Us1}-\ref{eq:Us4})]. 
Furthermore, they will enter at the dynamical level, by affecting {\em both} $P_{ee}$ and $P_{es}$, 
through the expressions of the mixing elements  $U_{ei}^m$ in matter, which  are determined by
the diagonalization of  the Hamiltonian 
%..........................................................................
\begin{equation}
    \bar H^{\,dyn}_{2\nu} = V_{CC}(x)
    \begin{pmatrix}
        \gamma^2 +  r_x\, |\alpha|^2 \  & r_x \, \alpha^* \beta\,  \\
        r_x\, \alpha\beta^*\,         & r_x\, |\beta|^2
    \end{pmatrix}
\,,    
\end{equation}
%..........................................................................
where the two parameters $\alpha$ and $\beta$ can be
complex numbers. With our choice of the phases,  $\beta = -s_{24}$ 
is still  real,  while $\alpha$ is complex and is obtained with 
the replacements ($s_{13} \to \tilde s_{13},  s_{14} \to \tilde s_{14})$ in 
its expression in Eq.~(\ref{eq:alfa_app}).

We postpone to a future work the study of the complex extension of the
treatment we have provided for the real case. Here we just limit ourselves  to observe 
that the CP phases always appear in terms involving two small mixing angles, 
and therefore it is difficult to observe their effects in current solar neutrino experiments.
Finally, we stress that for $\theta_{24} = \theta_{34} = 0$ the CP phases disappear
from the description of the solar neutrino transitions, thus rendering the numerical results 
presented in Sec.~III independent of them.

\bibliographystyle{h-physrev4}

\begin{thebibliography}{99}


% ------------------------------------------------------------------------------------------------------------
%						New reactor fluxes 
% ------------------------------------------------------------------------------------------------------------

%\cite{Mueller:2011nm}
\bibitem{Mueller:2011nm}
  T.~A.~Mueller {\it et al.},
  %``Improved Predictions of Reactor Antineutrino Spectra,''
  [arXiv:1101.2663 [hep-ex]].
  %%CITATION = ARXIV:1101.2663;%%

% ------------------------------------------------------------------------------------------------------------
%						Old reactor fluxes 
% ------------------------------------------------------------------------------------------------------------


\bibitem{Vogel:1980bk}
  P.~Vogel, G.~K.~Schenter, F.~M.~Mann, R.~E.~Schenter,
  %``Reactor Anti-neutrino Spectra And Their Application To Anti-neutrino Induced Reactions. 2.,''
  Phys.\ Rev.\  C {\bf 24}, 1543-1553 (1981).

\bibitem{VonFeilitzsch:1982jw}
  F.~Von Feilitzsch, A.~A.~Hahn, K.~Schreckenbach,
  %``Experimental Beta Spectra From Pu-239 And U-235 Thermal Neutron Fission Products And Their Correlated Anti-neutrinos Spectra,''
  Phys.\ Lett.\  B {\bf 118}, 162-166 (1982).
  
\bibitem{Schreckenbach:1985ep}
  K.~Schreckenbach, G.~Colvin, W.~Gelletly, F.~Von Feilitzsch,
  %``Determination Of The Anti-neutrino Spectrum From U-235 Thermal Neutron Fission Products Up To 9.5-mev,''
  Phys.\ Lett.\  B {\bf 160}, 325-330 (1985).

\bibitem{Hahn:1989zr}
  A.~A.~Hahn, K.~Schreckenbach, G.~Colvin, B.~Krusche, W.~Gelletly, F.~Von Feilitzsch,
  %``Anti-neutrino Spectra From Pu-241 And Pu-239 Thermal Neutron Fission Products,''
  Phys.\ Lett.\  B {\bf 218}, 365-368 (1989).
  



 % ------------------------------------------------------------------------------------------------------------
%				      The reactor antineutrino anomaly 
% ------------------------------------------------------------------------------------------------------------

 %\cite{Mention:2011rk}
\bibitem{Mention:2011rk}
  G.~Mention {\it et al.},
  %, M.~Fechner, T.~Lasserre, T.~A.~Mueller, D.~Lhuillier, M.~Cribier and A.~Letourneau,
  %``The Reactor Antineutrino Anomaly,''
  [arXiv:1101.2755 [hep-ex]].
  %%CITATION = ARXIV:1101.2755;%%
 
% ------------------------------------------------------------------------------------------------------------
%			 The calibration (Sage+Gallex) antineutrino anomaly 
% ------------------------------------------------------------------------------------------------------------
 
 %\cite{Abdurashitov:2005tb}
\bibitem{Abdurashitov:2005tb}
  J.~N.~Abdurashitov {\it et al.},
  %``Measurement of the response of a Ga solar neutrino experiment to neutrinos
  %from an Ar-37 source,''
  Phys.\ Rev.\  C {\bf 73}, 045805 (2006)
  [arXiv:nucl-ex/0512041].
  %%CITATION = PHRVA,C73,045805;%%

%\cite{Giunti:2010zu}
\bibitem{Giunti:2010zu}
  C.~Giunti, M.~Laveder,
  %``Statistical Significance of the Gallium Anomaly,''
  [arXiv:1006.3244 [hep-ph]].

% ------------------------------------------------------------------------------------------------------------
%			 ACCELERATORS anomalies 
% ------------------------------------------------------------------------------------------------------------


%\cite{Aguilar:2001ty}
\bibitem{Aguilar:2001ty}
  A.~Aguilar {\it et al.}  [LSND Collaboration],
  %``Evidence for neutrino oscillations from the observation of
  %anti-neutrino(electron) appearance in a anti-neutrino(muon) beam,''
  Phys.\ Rev.\  D {\bf 64}, 112007 (2001)
  [arXiv:hep-ex/0104049].
  %%CITATION = PHRVA,D64,112007;%%
  
  %\cite{AguilarArevalo:2007it}
\bibitem{AguilarArevalo:2007it}
  A.~A.~Aguilar-Arevalo {\it et al.}  [MiniBooNE Collaboration],
  %``A Search for electron neutrino appearance at the $\Delta m^{2} \sim
  %1$eV$^{2}$ scale,''
  Phys.\ Rev.\ Lett.\  {\bf 98}, 231801 (2007)
  [arXiv:0704.1500 [hep-ex]].
  %%CITATION = PRLTA,98,231801;%%
  
    %\cite{AguilarArevalo:2009xn}
\bibitem{AguilarArevalo:2009xn}
  A.~A.~Aguilar-Arevalo {\it et al.}  [MiniBooNE Collaboration],
  %``A Search for Electron Antineutrino Appearance at the Delta m**2 ~ 1-eV**2
  %Scale,''
  Phys.\ Rev.\ Lett.\  {\bf 103}, 111801 (2009)
  [arXiv:0904.1958 [hep-ex]].
  %%CITATION = PRLTA,103,111801;%%

  
% ------------------------------------------------------------------------------------------------------------
%	With new fluxes: Lesser Tension among appearance and disappearance
% ------------------------------------------------------------------------------------------------------------

 %\cite{Kopp:2011qd}
\bibitem{Kopp:2011qd}
  J.~Kopp, M.~Maltoni and T.~Schwetz,
  %``Are there sterile neutrinos at the eV scale?,''
  [arXiv:1103.4570 [hep-ph]].
  %%CITATION = ARXIV:1103.4570;%%
 

%\cite{Giunti:2010uj}
\bibitem{Giunti:2010uj}
  C.~Giunti, M.~Laveder,
  %``Large Short-Baseline antinu_mu Disappearance,''
  Phys.\ Rev.\  D\ {\bf 83}, 053006 (2011)
  [arXiv:1012.0267 [hep-ph]].
% [v3] Fri, 25 Mar 2011 16:19:00 GMT (81kb)

 
 % ------------------------------------------------------------------------------------------------------------
%	Nu_s and late cosmology
% ------------------------------------------------------------------------------------------------------------
 
 
 %\cite{Hamann:2010bk}
\bibitem{Hamann:2010bk}
  J.~Hamann, S.~Hannestad, G.~G.~Raffelt, I.~Tamborra and Y.~Y.~Y.~Wong,
  %``Cosmology seeking friendship with sterile neutrinos,''
  Phys.\ Rev.\ Lett.\  {\bf 105}, 181301 (2010)
  [arXiv:1006.5276 [hep-ph]].
  %%CITATION = PRLTA,105,181301;%%
 
 
 %\cite{Giusarma:2011ex}
\bibitem{Giusarma:2011ex}
  E.~Giusarma {\it et al.},
  % M.~Corsi, M.~Archidiacono, R.~de Putter, A.~Melchiorri, O.~Mena and S.~Pandolfi,
  %``Constraints on massive sterile neutrino species from current and future
  %cosmological data,''
  [arXiv:1102.4774 [astro-ph.CO]].
  %%CITATION = ARXIV:1102.4774;%%

%\cite{Hannestad:2005gj}
\bibitem{Hannestad:2005gj}
  S.~Hannestad,
  %``Neutrino masses and the dark energy equation of state - Relaxing the cosmological neutrino mass bound,''
  Phys.\ Rev.\ Lett.\  {\bf 95}, 221301 (2005)
  [astro-ph/0505551].

%\cite{Kristiansen:2011mp}
\bibitem{Kristiansen:2011mp}
  J.~R.~Kristiansen, O.~Elgaroy,
  %``Reactor sterile neutrinos, dark energy and the age of the universe,''
   [arXiv:1104.0704 [astro-ph.CO]].


% ------------------------------------------------------------------------------------------------------------
%	Nu_s and Nucleosynthesis
% ------------------------------------------------------------------------------------------------------------


%\cite{Izotov:2010ca}
\bibitem{Izotov:2010ca}
  Y.~I.~Izotov, T.~X.~Thuan,
  %``The primordial abundance of 4He: evidence for non-standard big bang nucleosynthesis,''
  Astrophys.\ J.\  {\bf 710}, L67-L71 (2010)
  [arXiv:1001.4440 [astro-ph.CO]].%\cite{Mangano:2011ar}

\bibitem{Mangano:2011ar}
  G.~Mangano and P.~D.~Serpico,
  %``A robust upper limit on $N_{\rm eff}$ from BBN, circa 2011,''
  arXiv:1103.1261 [astro-ph.CO].
  %%CITATION = ARXIV:1103.1261;%%

% ------------------------------------------------------------------------------------------------------------
%	solar Sterile ultralight 
% ------------------------------------------------------------------------------------------------------------
 
 %\cite{deHolanda:2003tx}
\bibitem{deHolanda:2003tx}
  P.~C.~de Holanda, A.~Yu.~Smirnov,
  %``Homestake result, sterile neutrinos and low-energy solar neutrino experiments,''
  Phys.\ Rev.\  D {\bf 69}, 113002 (2004)
  [hep-ph/0307266].
 
 %\cite{deHolanda:2010am}
\bibitem{deHolanda:2010am}
  P.~C.~de Holanda, A.~Yu.~Smirnov,
  %``Solar neutrino spectrum, sterile neutrinos and additional radiation in the Universe,''
   [arXiv:1012.5627 [hep-ph]].
 
% ------------------------------------------------------------------------------------------------------------
%	Nu_s at Nutel
% ------------------------------------------------------------------------------------------------------------

%\cite{Abbasi:2010ie}
\bibitem{Abbasi:2010ie}
  R.~Abbasi {\it et al.}  [IceCube Collaboration],
  %``Measurement of the atmospheric neutrino energy spectrum from 100 GeV to 400
  %TeV with IceCube,''
  Phys.\ Rev.\  D {\bf 83} (2011) 012001
  [arXiv:1010.3980 [astro-ph.HE]].
  %%CITATION = PHRVA,D83,012001;%%

%\cite{Nunokawa:2003ep}
\bibitem{Nunokawa:2003ep}
  H.~Nunokawa, O.~L.~G.~Peres, R.~Zukanovich Funchal,
  %``Probing the LSND mass scale and four neutrino scenarios with a neutrino telescope,''
  Phys.\ Lett.\  B {\bf 562}, 279-290 (2003)
  [hep-ph/0302039].


\bibitem{Choubey:2007ji}
  S.~Choubey,
  %``Signature of sterile species in atmospheric neutrino data at neutrino telescopes,''
  JHEP {\bf 0712}, 014 (2007)
  [arXiv:0709.1937 [hep-ph]].
 
 %\cite{Razzaque:2011ab}
\bibitem{Razzaque:2011ab}
  S.~Razzaque, A.~Yu.~Smirnov,
  %``Searching for sterile neutrinos in ice,''
[arXiv:1104.1390 [hep-ph]].


% ------------------------------------------------------------------------------------------------------------
%	Solar sterile as a leading mechansim
% ------------------------------------------------------------------------------------------------------------


\bibitem{Barger:1990bg}
  V.~D.~Barger, N.~Deshpande, P.~B.~Pal, R.~J.~N.~Phillips, K.~Whisnant,
  %``Sterile-neutrino solutions to the solar puzzle,''
  Phys.\ Rev.\  D {\bf 43}, 1759-1762 (1991).
  
\bibitem{Krastev:1996gc}
  P.~I.~Krastev, S.~T.~Petcov, L.~Qiuyu,
  %``On the MSW electron-neutrino ---> sterile neutrino transition solution of the solar neutrino problem,''
  Phys.\ Rev.\  {\bf D54}, 7057-7066 (1996)
  [hep-ph/9602333].

% ------------------------------------------------------------------------------------------------------------
%	Solar sterile subleading analyses
% ------------------------------------------------------------------------------------------------------------


\bibitem{Giunti:2000wt}
  C.~Giunti, M.~C.~Gonzalez-Garcia, C.~Pena-Garay,
  %``Four-neutrino oscillation solutions of the solar neutrino problem,''
  Phys.\ Rev.\  D {\bf 62}, 013005 (2000)
  [hep-ph/0001101].

 \bibitem{Bahcall:2002zh}
  J.~N.~Bahcall, M.~C.~Gonzalez-Garcia and C.~Pena-Garay,
  %``If sterile neutrinos exist, how can one determine the total neutrino
  %fluxes?,''
  Phys.\ Rev.\  C {\bf 66}, 035802 (2002)
  [arXiv:hep-ph/0204194].
  %%CITATION = PHRVA,C66,035802;%%
 
% + sterile ultralight + ue3
  
%\cite{deHolanda:2002ma}
\bibitem{deHolanda:2002ma}
  P.~C.~de Holanda and A.~Yu.~Smirnov,
  %``Searches for sterile component with solar neutrinos and KamLAND,''
  [arXiv:hep-ph/0211264].
  %%CITATION = HEP-PH/0211264;%%
 
 %\cite{Bahcall:2002ij}
\bibitem{Bahcall:2002ij}
  J.~N.~Bahcall, M.~C.~Gonzalez-Garcia, C.~Pena-Garay,
  %``Solar neutrinos before and after KamLAND,''
  JHEP {\bf 0302}, 009 (2003)
  [hep-ph/0212147].
 
  %\cite{Maltoni:2002ni}
\bibitem{Maltoni:2002ni}
  M.~Maltoni, T.~Schwetz, M.~A.~Tortola, J.~W.~F.~Valle,
  %``Constraining neutrino oscillation parameters with current solar and atmospheric data,''
  Phys.\ Rev.\  D {\bf 67}, 013011 (2003)
  [hep-ph/0207227].
 
 %\cite{Cirelli:2004cz}
\bibitem{Cirelli:2004cz}
  M.~Cirelli, G.~Marandella, A.~Strumia, F.~Vissani,
  %``Probing oscillations into sterile neutrinos with cosmology, astrophysics and experiments,''
  Nucl.\ Phys.\  B {\bf 708}, 215-267 (2005)
  [hep-ph/0403158].
 
 %\cite{GonzalezGarcia:2007ib}
\bibitem{GonzalezGarcia:2007ib}
  M.~C.~Gonzalez-Garcia and M.~Maltoni,
  %``Phenomenology with Massive Neutrinos,''
  Phys.\ Rept.\  {\bf 460}, 1 (2008)
  [arXiv:0704.1800 [hep-ph]].
  %%CITATION = PRPLC,460,1;%%


% ------------------------------------------------------------------------------------------------------------
%	Solar 2+2 theory
% ------------------------------------------------------------------------------------------------------------

%\cite{Dooling:1999sg}
\bibitem{Dooling:1999sg}
  D.~Dooling, C.~Giunti, K.~Kang, C.~W.~Kim,
  %``Matter effects in four neutrino mixing,''
  Phys.\ Rev.\  D {\bf 61}, 073011 (2000)
  [hep-ph/9908513].

% ------------------------------------------------------------------------------------------------------------
%	CHOOZ BOUND on theta_13
% ------------------------------------------------------------------------------------------------------------

\bibitem{Apollonio:2002gd}
M.~Apollonio {\em et~al.} (CHOOZ Collaboration),
\newblock Eur. Phys. J. C {\bf 27}, 331 (2003).
%%

% ------------------------------------------------------------------------------------------------------------
%	BUGEY BOUND on theta_14
% ------------------------------------------------------------------------------------------------------------

%\cite{Declais:1994su}
\bibitem{Declais:1994su}
  Y.~Declais {\it et al.},
  %``Search for neutrino oscillations at 15-meters, 40-meters, and 95-meters
  %from a nuclear power reactor at Bugey,''
  Nucl.\ Phys.\  B {\bf 434}, 503 (1995).
  %%CITATION = NUPHA,B434,503;%%


% ------------------------------------------------------------------------------------------------------------
%	HINT of theta_13>0
% ------------------------------------------------------------------------------------------------------------

%\cite{Fogli:2008jx}
\bibitem{Fogli:2008jx}
  G.~L.~Fogli, E.~Lisi, A.~Marrone, A.~Palazzo and A.~M.~Rotunno,
  %``Hints of theta(13) > 0 from global neutrino data analysis,''
  Phys.\ Rev.\ Lett.\  {\bf 101}, 141801 (2008)
  [arXiv:0806.2649 [hep-ph]].
  %%CITATION = PRLTA,101,141801;%%
  
% ------------------------------------------------------------------------------------------------------------
%	HINT of NSI
% ------------------------------------------------------------------------------------------------------------

\bibitem{Palazzo:2011vg}
  A.~Palazzo,
  %``Hint of non-standard Mikheyev-Smirnov-Wolfenstein dynamics in solar neutrino conversion,''
  Phys.\ Rev.\  D {\bf 83}, 101701(R) (2011)
  [arXiv:1101.3875 [hep-ph]].
  %%CITATION = ARXIV:1101.3875;%%

\bibitem{Palazzo:2009rb}
  A.~Palazzo and J.~W.~F.~Valle,
  %``Confusing non-zero theta(13) with non-standard interactions in the solar neutrino sector,''
  Phys.\ Rev.\  D {\bf 80}, 091301(R) (2009)
  [arXiv:0909.1535 [hep-ph]].


% ------------------------------------------------------------------------------------------------------------
%	solar 3+1 theory
% ------------------------------------------------------------------------------------------------------------

 %\cite{Giunti:2009xz}
\bibitem{Giunti:2009xz}
  C.~Giunti and Y.~F.~Li,
  %``Matter Effects in Active-Sterile Solar Neutrino Oscillations,''
  Phys.\ Rev.\  D {\bf 80}, 113007 (2009)
  [arXiv:0910.5856 [hep-ph]].
  %%CITATION = PHRVA,D80,113007;%%


% ------------------------------------------------------------------------------------------------------------
%	number of parameters of the lepton Mixing Matrix
% ------------------------------------------------------------------------------------------------------------

%\cite{Schechter:1980gr}
\bibitem{Schechter:1980gr}
  J.~Schechter, J.~W.~F.~Valle,
  %``Neutrino Masses in SU(2) x U(1) Theories,''
  Phys.\ Rev.\  D {\bf 22}, 2227 (1980).
  
% ------------------------------------------------------------------------------------------------------------
%	Parametrization of the lepton Mixing Matrix used by PDG
% ------------------------------------------------------------------------------------------------------------
 
 %\cite{PDG}
\bibitem{PDG}
K. Nakamura et al. (Particle Data Group),
J. Phys. G 37, 075021 (2010).   
% ------------------------------------------------------------------------------------------------------------
%	Parametrization of the lepton Mixing Matrix used for ATM and LBL
% ------------------------------------------------------------------------------------------------------------
  
  %\cite{Maltoni:2007zf}
\bibitem{Maltoni:2007zf}
  M.~Maltoni, T.~Schwetz,
  %``Sterile neutrino oscillations after first MiniBooNE results,''
  Phys.\ Rev.\  D {\bf 76}, 093005 (2007).
%  [arXiv:0705.0107 [hep-ph]].
%\cite{Adamson:2010wi}

\bibitem{Adamson:2010wi}
  P.~Adamson {\it et al.}  [The MINOS Collaboration],
  %``Search for sterile neutrino mixing in the MINOS long baseline experiment,''
  Phys.\ Rev.\  D {\bf 81}, 052004 (2010)
  [arXiv:1001.0336 [hep-ex]].
  %%CITATION = PHRVA,D81,052004;%%



% ------------------------------------------------------------------------------------------------------------
%	MSW
% ------------------------------------------------------------------------------------------------------------

\bibitem{Wolfenstein:1977ue}
L.~Wolfenstein,
\newblock Phys. Rev. D {\bf 17}, 2369 (1978).
%%CITATION = PHRVA,D17,2369;%%

\bibitem{smirnov}  
                S.~P.~Mikheev and A.~Yu.\ Smirnov,
                Yad.\ Fiz.\ {\bf 42}, 1441 (1985)
                [Sov.\ J.\ Nucl.\ Phys.\ {\bf 42}, 913 (1985)].

% ------------------------------------------------------------------------------------------------------------
%	Pee formula 2nu adiabatic  &	Reduction to 2x2 dynamics 
% ------------------------------------------------------------------------------------------------------------

%\cite{Kuo:1989qe}
\bibitem{Kuo:1989qe}
  T.~K.~Kuo and J.~T.~Pantaleone,
  %``Neutrino Oscillations in Matter,''
  Rev.\ Mod.\ Phys.\  {\bf 61}, 937 (1989).
  %%CITATION = RMPHA,61,937;%%
  

  
% ------------------------------------------------------------------------------------------------------------
%	c^2_13 factor in 3nu solar
% ------------------------------------------------------------------------------------------------------------

%\cite{Shi:1991zw}
\bibitem{Shi:1991zw}
  X.~Shi and D.~N.~Schramm,
  %``Solar neutrinos and the MSW effect for three neutrino mixing,''
  Phys.\ Lett.\  B {\bf 283}, 305 (1992).
 
\bibitem{Fogli:1993ck}
  G.~L.~Fogli, E.~Lisi and D.~Montanino,
  %``A comprehensive analysis of solar, atmospheric, accelerator and reactor
  %neutrino experiments in a hierarchical three generation scheme,''
  Phys.\ Rev.\  D {\bf 49}, 3626 (1994).
  %%CITATION = PHRVA,D49,3626;%%}

% ------------------------------------------------------------------------------------------------------------
%	Mild energy dependence induced by dynamic effect V -> V c^2_13 in 3nu solar
% ------------------------------------------------------------------------------------------------------------
   
 \bibitem{Fogli:2005cq}
  G.~L.~Fogli, E.~Lisi, A.~Marrone and A.~Palazzo,
  %``Global analysis of three-flavor neutrino masses and mixings,''
  Prog.\ Part.\ Nucl.\ Phys.\  {\bf 57}, 742 (2006)
  [arXiv:hep-ph/0506083].
  %%CITATION = PPNPD,57,742;%%}   

% ------------------------------------------------------------------------------------------------------------
%	Solar Data
% ------------------------------------------------------------------------------------------------------------

\bibitem{cleveland:1998nv}
B.~T. Cleveland {\em et~al.},
\newblock Astrophys. J. {\bf 496}, 505 (1998).
%%CITATION = ASJOA,496,505;%%

\bibitem{abdurashitov:2002nt}
J.~N. Abdurashitov {\em et~al.} (SAGE Collaboration),
\newblock J. Exp. Theor. Phys. {\bf 95}, 181 (2002).
%%CITATION = ASTRO-PH 0204245;%%

\bibitem{Hampel:1998xg}
W.~Hampel {\em et~al.} (GALLEX Collaboration),
\newblock Phys. Lett. B {\bf 447}, 127 (1999).
%%CITATION = PHLTA,B447,127;%%

\bibitem{Altmann:2005ix}
M.~Altmann {\em et~al.} (GNO Collaboration),
\newblock Phys. Lett. B {\bf 616}, 174 (2005).
%%CITATION = HEP-EX/0504037;%%

\bibitem{kirsten2008retrospect}
T.~Kirsten,
\newblock  J. Phys. Conf. Ser {\bf120},  052013 (2008).


\bibitem{fukuda}
S.~Fukuda {\em et~al.} (Super-Kamiokande Collaboration),
\newblock Phys. Rev. Lett. {\bf 86}, 5651 (2001); {\bf 86}, 5656 (2001); Phys. Lett. B {\bf 539}, 179 (2002).


\bibitem{ahmad}
Q.~R. Ahmad {\em et~al.} (SNO Collaboration),
\newblock Phys. Rev. Lett. {\bf 87}, 071301 (2001);  {\bf 89}, 011301 (2002);  {\bf 89}, 011302 (2002).

\bibitem{Ahmed:2003kj}
S.~N. Ahmed {\em et~al.} (SNO Collaboration),
\newblock Phys. Rev. Lett. {\bf 92}, 181301 (2004).
%%CITATION = NUCL-EX/0309004;%%

\bibitem{Aharmim:2005gt}
B.~Aharmim {\em et~al.} (SNO Collaboration), 
\newblock Phys. Rev. C {\bf 72}, 055502 (2005).
%%CITATION = NUCL-EX 0502021;%%

\bibitem{Aharmim:2008kc}
B.~Aharmim {\em et~al.} (SNO Collaboration), 
\newblock Phys. Rev. Lett. {\bf 101}, 111301 (2008).
%%CITATION = 0806.0989;%%


\bibitem{Collaboration:2011rx}
  G.~Bellini {\it et al.} (Borexino Collaboration),
  %``Precision measurement of the 7Be solar neutrino interaction rate in
  %Borexino,''
  [arXiv:1104.1816 [hep-ex]].
  %%CITATION = ARXIV:1104.1816;%%\bibitem{Arpesella:2008mt}


%\cite{Bellini:2008mr}
\bibitem{Bellini:2008mr}
  G.~Bellini {\it et al.}  (Borexino Collaboration),
  %``Measurement of the solar 8B neutrino rate with a liquid scintillator target
  %and 3 MeV energy threshold in the Borexino detector,''
  Phys.\ Rev.\  D {\bf 82}, 033006 (2010).
%  [arXiv:0808.2868 [astro-ph]].
  %%CITATION = PHRVA,D82,033006;%%

\bibitem{Aharmim:2009gd}
  B.~Aharmim {\it et al.}  (SNO Collaboration),
%  ``Low Energy Threshold Analysis of the Phase I and Phase II Data Sets of the
 % Sudbury Neutrino Observatory,''
  Phys.\ Rev.\  C {\bf 81}, 055504 (2010).
 % [arXiv:0910.2984 [nucl-ex]].
  %%CITATION = PHRVA,C81,055504;%% (LETA),

%\cite{Gando:2010aa}
\bibitem{Gando:2010aa}
  A.~Gando {\it et al.} (KamLAND Collaboration),
  %``Constraints on $\theta_{13}$ from A Three-Flavor Oscillation Analysis of Reactor Antineutrinos at KamLAND,''
  Phys.\ Rev.\  D {\bf 83}, 052002 (2011)
  [arXiv:1009.4771 [hep-ex]].

% ------------------------------------------------------------------------------------------------------------
%	other 3nu solar analyses
% ------------------------------------------------------------------------------------------------------------

\bibitem{Balantekin:2008zm}
A.~B. Balantekin and D.~Yilmaz,
\newblock J. Phys. G {\bf 35}, 075007 (2008).
%%CITATION = 0804.3345;%%

\bibitem{Schwetz:2008er}
T.~Schwetz, M.~Tortola and J.~W.~F. Valle,
\newblock New J. Phys. {\bf 10}, 113011 (2008).
%%CITATION = 0808.2016;%%

\bibitem{Ge:2008sj}
H.~L.~Ge, C.~Giunti and Q.~Y.~Liu,
\newblock Phys. Rev. D {\bf 80}, 053009 (2009).
%%CITATION = 0810.5443;%%

\bibitem{GonzalezGarcia:2010er}
  M.~C.~Gonzalez-Garcia, M.~Maltoni and J.~Salvado,
  %``Updated global fit to three neutrino mixing: status of the hints of theta13
  %> 0,''
  JHEP {\bf 1004}, 056 (2010)
  [arXiv:1001.4524 [hep-ph]].
  %%CITATION = JHEPA,1004,056;%%

%\cite{Schwetz:2011qt}
\bibitem{Schwetz:2011qt}
  T.~Schwetz, M.~Tortola, J.~W.~F.~Valle,
  %``Global neutrino data and recent reactor fluxes: status of three-flavour oscillation parameters,''
  [arXiv:1103.0734 [hep-ph]].


% ------------------------------------------------------------------------------------------------------------
%	response functions
% ------------------------------------------------------------------------------------------------------------


%\cite{Villante:1998pe}
\bibitem{Villante:1998pe}
  F.~L.~Villante, G.~Fiorentini and E.~Lisi,
  %``Solar neutrino interactions: Using charged currents at SNO to tell neutral
  %currents at Super-Kamiokande,''
  Phys.\ Rev.\  D {\bf 59} (1999) 013006
  [arXiv:hep-ph/9807360].
  %%CITATION = PHRVA,D59,013006;%%

%\cite{Fogli:2001nn}
\bibitem{Fogli:2001nn}
  G.~L.~Fogli, E.~Lisi, A.~Palazzo and F.~L.~Villante,
  %``Solar neutrino event spectra: Tuning SNO to equalize Super-Kamiokande,''
  Phys.\ Rev.\  D {\bf 63}, 113016 (2001)
  [arXiv:hep-ph/0102288].
  %%CITATION = PHRVA,D63,113016;%%

% ------------------------------------------------------------------------------------------------------------
%	Theta_13 sensisitivity of solar (LE HE synergy)
% ------------------------------------------------------------------------------------------------------------

\bibitem{Goswami:2004cn}
  S.~Goswami and A.~Yu.~Smirnov,
  %``Solar neutrinos and 1-3 leptonic mixing,''
  Phys.\ Rev.\  D {\bf 72}, 053011 (2005)
  [arXiv:hep-ph/0411359].
  %%CITATION = PHRVA,D72,053011;%%

\bibitem{Fogli:2006fu}
  G.~L.~Fogli, E.~Lisi, A.~Marrone and A.~Palazzo,
  Proceedings of the
International Workshop: Neutrino Oscillations\\
ed. by M. Baldo Ceolin, Univ. of Padova publication (2006), Vol.
I, p. 69.
  %``Solar neutrinos: With a tribute to John. N. Bahcall,''
  [arXiv:hep-ph/0605186].
  %%CITATION = HEP-PH/0605186;%%}, due to the different dependence on the two mixing angles 
  
% ------------------------------------------------------------------------------------------------------------
%	(1+eps cot theta) = 0 condition
% ------------------------------------------------------------------------------------------------------------

\bibitem{Bergmann:1997mr}
  S.~Bergmann,
  %``The Solar neutrino problem in the presence of flavor changing neutrino
  %interactions,''
  Nucl.\ Phys.\  B {\bf 515}, 363 (1998)
  [arXiv:hep-ph/9707398].
  %%CITATION = NUPHA,B515,363;%% non-standard neutrino interaction of the flavor

% ------------------------------------------------------------------------------------------------------------
%	average of phases 
% ------------------------------------------------------------------------------------------------------------

%\cite{Lisi:2000su}
\bibitem{Lisi:2000su}
  E.~Lisi, A.~Marrone, D.~Montanino, A.~Palazzo, S.~T.~Petcov,
  %``Analytical description of quasivacuum oscillations of solar neutrinos,''
  Phys.\ Rev.\  D {\bf 63}, 093002 (2001)
  [hep-ph/0011306].

% ------------------------------------------------------------------------------------------------------------
%	CP violation appendix
% ------------------------------------------------------------------------------------------------------------

\bibitem{Kuo:1987km}
  T.~K.~Kuo and J.~T.~Pantaleone,
  %``T NONCONSERVATION IN THREE NEUTRINO OSCILLATIONS,''
  Phys.\ Lett.\  B {\bf 198}, 406 (1987).
  %%CITATION = PHLTA,B198,406;%%

%\cite{Minakata:1999ze}
\bibitem{Minakata:1999ze}
  H.~Minakata and S.~Watanabe,
  %``Solar neutrinos and leptonic CP violation,''
  Phys.\ Lett.\  B {\bf 468}, 256 (1999)
  [arXiv:hep-ph/9906530].
  %%CITATION = PHLTA,B468,256;%%

%\cite{Yokomakura:2002av}
\bibitem{Yokomakura:2002av}
  H.~Yokomakura, K.~Kimura and A.~Takamura,
  %``Overall feature of CP dependence for neutrino oscillation probability in
  %arbitrary matter profile,''
  Phys.\ Lett.\  B {\bf 544}, 286 (2002)
  [arXiv:hep-ph/0207174].
  %%CITATION = PHLTA,B544,286;%%
  
%\cite{Balantekin:2007es}
\bibitem{Balantekin:2007es}
  A.~B.~Balantekin, J.~Gava, C.~Volpe,
  %``Possible CP-Violation effects in core-collapse Supernovae,''
  Phys.\ Lett.\  B {\bf 662}, 396-404 (2008)
  [arXiv:0710.3112 [astro-ph]].
  





\end{thebibliography}

\end{document}